\documentclass[twoside,twocolumn,9pt]{article}
\usepackage{extsizes}
\usepackage[super,sort&compress,comma]{natbib}
\usepackage[version=3]{mhchem}
\usepackage[left=1.5cm, right=1.5cm, top=1.785cm, bottom=2.0cm]{geometry}
\usepackage{balance}
\usepackage{mathptmx}
\usepackage{sectsty}
\usepackage{graphicx}
\usepackage{lastpage} 
\usepackage[format=plain,justification=justified, singlelinecheck=false,font={stretch=1.125,small,sf},labelfont=bf,labelsep=space]{caption} 
\usepackage{float}
\usepackage{fancyhdr}
\usepackage{fnpos}
\usepackage[english]{babel}

\usepackage{array}
\usepackage{droidsans}
\usepackage{charter}
\usepackage[T1]{fontenc}
\usepackage[usenames,dvipsnames]{xcolor}
\usepackage{setspace}
\usepackage[compact]{titlesec}
\usepackage{hyperref}
\usepackage{physics}
\usepackage{nicefrac}
\usepackage{subcaption}
\usepackage{chemformula}
\usepackage{xspace}
\usepackage{nicefrac}
\usepackage{ulem} %To cross out text

\usepackage{amsmath}
\usepackage{amssymb}
\usepackage{amscd}
\usepackage[np,autolanguage]{numprint}
\usepackage{bm}
\usepackage{xcolor}

\usepackage{hyperref}
\usepackage{cleveref}    % For \cref
\usepackage{braket}      % For \Braket
\usepackage{upgreek}     % For \uppi
\usepackage{chemformula} % For \ch{} 
\usepackage{booktabs}    % Better \toprule \midrule \bottomrule
\definecolor{cream}{RGB}{222,217,201}

\begin{document}

\pagestyle{fancy}
\thispagestyle{plain}
\fancypagestyle{plain}{
%%%HEADER%%%
\renewcommand{\headrulewidth}{0pt}
}
%%%END OF HEADER%%%

%%%PAGE SETUP - Please do not change any commands within this section%%%
\makeFNbottom
\makeatletter
\renewcommand\LARGE{\@setfontsize\LARGE{15pt}{17}}
\renewcommand\Large{\@setfontsize\Large{12pt}{14}}
\renewcommand\large{\@setfontsize\large{10pt}{12}}
\renewcommand\footnotesize{\@setfontsize\footnotesize{7pt}{10}}
\makeatother

\renewcommand{\thefootnote}{\fnsymbol{footnote}}
\renewcommand\footnoterule{\vspace*{1pt}%
\color{cream}\hrule width 3.5in height 0.4pt \color{black}\vspace*{5pt}}
\setcounter{secnumdepth}{5}

\makeatletter
\renewcommand\@biblabel[1]{#1}
\renewcommand\@makefntext[1]%
{\noindent\makebox[0pt][r]{\@thefnmark\,}#1}
\makeatother
\renewcommand{\figurename}{\small{Fig.}~}
\sectionfont{\sffamily\Large}
\subsectionfont{\normalsize}
\subsubsectionfont{\bf}
\setstretch{1.125} %In particular, please do not alter this line.
\setlength{\skip\footins}{0.8cm}
\setlength{\footnotesep}{0.25cm}
\setlength{\jot}{10pt}
\titlespacing*{\section}{0pt}{4pt}{4pt}
\titlespacing*{\subsection}{0pt}{15pt}{1pt}
%%%END OF PAGE SETUP%%%

%%%FOOTER%%%
\fancyfoot{}
\fancyfoot[LO,RE]{\vspace{-7.1pt}\includegraphics[height=9pt]{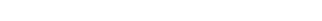}}
\fancyfoot[CO]{\vspace{-7.1pt}\hspace{11.9cm}\includegraphics{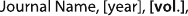}}
\fancyfoot[CE]{\vspace{-7.2pt}\hspace{-13.2cm}\includegraphics{head_foot/RF}}
\fancyfoot[RO]{\footnotesize{\sffamily{1--\pageref{LastPage} ~\textbar  \hspace{2pt}\thepage}}
}
\fancyfoot[LE]{\footnotesize{\sffamily{\thepage~\textbar\hspace{4.65cm} 1--\pageref{LastPage}}
}}
\fancyhead{}
\renewcommand{\headrulewidth}{0pt}
\renewcommand{\footrulewidth}{0pt}
\setlength{\arrayrulewidth}{1pt}
\setlength{\columnsep}{6.5mm}
\setlength\bibsep{1pt}
%%%END OF FOOTER%%%

%%%FIGURE SETUP - please do not change any commands within this section%%%
\makeatletter
\newlength{\figrulesep}
\setlength{\figrulesep}{0.5\textfloatsep}

\newcommand{\topfigrule}{\vspace*{-1pt}%
\noindent{\color{cream}\rule[-\figrulesep]{\columnwidth}{1.5pt}} }

\newcommand{\botfigrule}{\vspace*{-2pt}%
\noindent{\color{cream}\rule[\figrulesep]{\columnwidth}{1.5pt}} }

\newcommand{\dblfigrule}{\vspace*{-1pt}%
\noindent{\color{cream}\rule[-\figrulesep]{\textwidth}{1.5pt}} }

\makeatother
%%%END OF FIGURE SETUP%%%

%%%TITLE, AUTHORS AND ABSTRACT%%%
\twocolumn[
  \begin{@twocolumnfalse}
{\includegraphics[height=30pt]{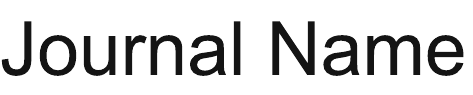}\hfill\raisebox{0pt}[0pt][0pt]{\includegraphics
[height=55pt]{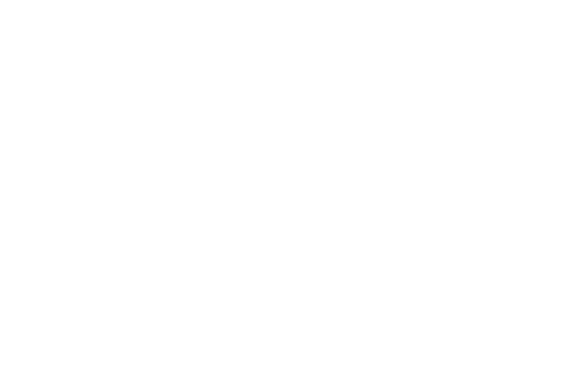}}\\[1ex]
\includegraphics[width=18.5cm]{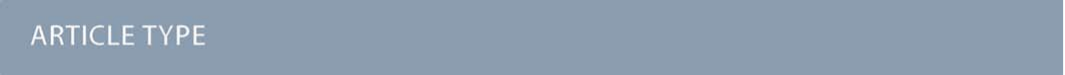}}\par
\vspace{1em}
\sffamily
\begin{tabular}{m{4.5cm} p{13.5cm} }

\includegraphics{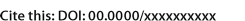} & \noindent\LARGE{\textbf{Consistent GMTKN55 and molecular-crystal accuracy using minimally empirical DFT with XDM(Z) dispersion$^\dag$}} \\
\vspace{0.3cm} & \vspace{0.3cm} \\

%\title{The exchange-correlation dipole moment dispersion method}

 & \noindent\large{Kyle R. Bryenton\textit{$^{a,b}$} and Erin R.\ Johnson\textit{$^{a,b,c\ast}$}} \\

%\author{Kyle R. Bryenton}
%\affiliation{
%Department of Physics and Atmospheric Science, Dalhousie University, 
%6310 Coburg Road, Halifax, Nova Scotia, Canada, B3H 4R2}

%\author{Erin R. Johnson}
%\email{erin.johnson@dal.ca}
%\affiliation{
%Department of Physics and Atmospheric Science, Dalhousie University, 
%6310 Coburg Road, Halifax, Nova Scotia, Canada, B3H 4R2}
%\affiliation{
%Department of Chemistry, Dalhousie University, 6243 Alumni Crescent, Halifax, Nova Scotia, Canada, B3H 4R2}
%\affiliation{
%Yusuf Hamied Department of Chemistry, University of Cambridge
%Lensfield Road, Cambridge, UK, CB2 1EW}
    
%\date{\today}

%\begin{abstract}

\includegraphics{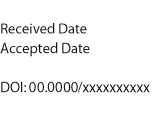} & \noindent\normalsize{%
Density-functional theory (DFT) has become the workhorse of modern computational chemistry, with dispersion corrections such as the exchange-hole dipole moment (XDM) model playing a key role in high-accuracy modelling of large-scale systems. All previous production implementations of XDM have used the two-parameter Becke--Johnson damping function based on atomic radii. Here, we introduce and implement a new XDM variant that uses a one-parameter damping function based on atomic numbers, recently proposed by Becke. Both this new Z damping and the canonical BJ-damping variants of XDM are benchmarked on the comprehensive GMTKN55 database using minimally empirical generalised-gradient-approximation, global hybrid, and range-separated hybrid functionals. This marks the first time that the XDM (and many-body dispersion, MBD) corrections have been tested on the GMTKN55 set. Using the new WTMAD-4 metric, an outlier analysis is performed for all new data, as well as for top-ranking functionals from the literature at each rung, providing insight into both performance and consistency across the dataset. We also extended our analysis to the DM21 and Skala machine-learned functionals that have garnered recent attention. To test Z damping's transferability to the solid state, four benchmarks involving molecular crystals are also considered. Across these molecular and solid-state benchmarks, the revPBE0 and B86bPBE0 hybrid functionals, paired with the Z-damped XDM variant, show excellent performance.

%\end{abstract}
} \\%The abstract goes here instead of the text "The abstract should be..."

\end{tabular}

 \end{@twocolumnfalse} \vspace{0.6cm}

  ]
%%%END OF TITLE, AUTHORS AND ABSTRACT%%%

%%%FONT SETUP - please do not change any commands within this section
\renewcommand*\rmdefault{bch}\normalfont\upshape
\rmfamily
\section*{}
\vspace{-1cm}

%%%FOOTNOTES%%%

\footnotetext{\textit{$^{a}$~
Department of Physics and Atmospheric Science, Dalhousie University, 6310 Coburg Road, Halifax, Nova Scotia, Canada, B3H 4R2}}

\footnotetext{\textit{$^{b}$~Department of Chemistry, Dalhousie University, 6243 Alumni Crescent, Halifax, Nova Scotia, B3H 4R2, Canada. E-mail: erin.johnson@dal.ca}}

\footnotetext{\textit{$^{c}$~Yusuf Hamied Department of Chemistry, University of Cambridge, Lensfield Road, Cambridge, CB2 1EW, United Kingdom.}}

\footnotetext{\dag~Electronic Supplementary Information (ESI) available, see DOI: 10.1039/cXCP00000x/}

%%%END OF FOOTNOTES%%%

%%%MAIN TEXT%%%%

%\maketitle

\section{Introduction}

Despite being the weakest of the van der Waals forces, London dispersion interactions are collectively extremely important in determining the structural and energetic properties of many chemical systems. Because dispersion physics is not included in most density-functional approximations (DFAs) for modelling electronic structure, they are commonly augmented by a dispersion correction (DC). Numerous such dispersion methods exist in the literature and may be divided into two classes: (i) explicitly non-local corrections that are included within the self-consistent field (SCF) procedure; and (ii) additive corrections, which can be simply geometry based, or dependent on the SCF electron density. The first type includes the family of van der Waals functionals (vdW-DF),\cite{dion2004van, roman2009efficient, lee2010higher} as well as (r)VV10.\cite{vydrov2010nonlocal,sabatini2013nonlocal} However, due to their non-local nature, these methods are significantly more expensive than additive corrections. Popular additive corrections include the Grimme-D series (D1,\cite{grimme2004accurate} D2,\cite{grimme2006semiempirical} D3(0),\cite{grimme2010consistent} D3(BJ),\cite{grimme2011effect}  D4\cite{caldeweyher2019generally}); the many-body dispersion family (TS,\cite{tkatchenko2009accurate} MBD@rsSCS,\cite{tkatchenko2012accurate, ambrosetti2014long} \mbox{MBD-NL}\cite{hermann2020density} uMBD,\cite{kim2020umbd} MBD-FI\cite{gould2016fractionally}); and the exchange-hole dipole moment (XDM) model.\cite{johnson2017exchange}

XDM was originally formulated between 2005 and 2007\cite{becke2005density,johnson2006post,becke2007exchange} and has since proven to be one of the most broadly accurate DFA dispersion treatments due to its limited empiricism and inclusion of important physical considerations.\cite{bryenton2023many} XDM has demonstrated accuracy, efficiency, and stability in modelling dispersion binding across a highly diverse range of chemical systems, including intermolecular complexes,\cite{otero2013non,nickerson2023comparison} bulk metals,\cite{adeleke2023effects} salts,\cite{otero2020application,christian2021interplay} layered materials,\cite{otero2020asymptotic} surfaces,\cite{christian2016surface,christian2017adsorption2} and molecular crystals.\cite{otero2012benchmark,otero2014predicting} The recent implementation of XDM in the FHI-aims\cite{blum2009ab} software, and pairing with hybrid functionals, allows computation of molecular crystal lattice energies with the highest accuracy of any dispersion-corrected DFT reported to date.\cite{price2023xdm} It has also shown great success in the area of molecular crystal structure prediction (CSP).\cite{price2023accurate,mayo2024assessment} 

However, Becke recently showed\cite{becke2024remarkably} that XDM fails to accurately predict the binding energies of two alkali-metal clusters (Li$_8$ and Na$_8$) in the ALK8 subset of the GMTKN55 thermochemistry benchmark.\cite{goerigk2017look} The error was traced to the Becke--Johnson (BJ) damping function\cite{johnson2006post} used in XDM to damp the dispersion energy to a small negative value at short interatomic separations. An alternative damping function based on atomic numbers, $Z$, was proposed and found to provide good accuracy for these metal clusters, and the GMTKN55 benchmark as a whole.\cite{becke2024remarkably} Notably, the $Z$-dependent damping function is simpler, relying on only one empirical parameter for use with a given DFA, as opposed to the two parameters used in BJ damping. However, the performance of Z damping has not yet been assessed on solid-state systems, or in conjunction with any other density functionals beyond the DH24 double hybrid. For the present study, Z damping was implemented in the FHI-aims code, and the performance of BJ- and Z-damped variants of XDM, paired with an assortment of minimally empirical density functionals, is assessed for the GMTKN55 and selected molecular-crystal benchmarks. Further, the Z-damping function has been implemented in the open-source code, PostG,\cite{otero2013non, otero2025postgxcdm} which allows the XDM(Z) dispersion correction to be applied \textit{ad hoc} to any of the dozens of quantum-chemical codes that write  \texttt{.wfx}, \texttt{.wfn}, or \texttt{.molden}
files.\footnote{Since many dialects of the \texttt{.molden} filetype exist, users must verify the input file is handled correctly.}
%Overall, ...

\section{Theory}

The XDM dispersion energy is written as a sum over all pairs of atoms, $i$ and $j$:
\begin{equation}
E_\text{disp}^\text{XDM} = - \sum_{i<j} \left( \frac{C_{6,ij} f_6}{R^6_{ij}} + \frac{C_{8,ij} f_8}{R^8_{ij}} + \frac{C_{10,ij} f_{10}}{R^{10}_{ij}} \right)\,.
\end{equation}
Here, $C_n$ dispersion coefficients are computed for each atom pair from the self-consistent electron density of the system, as well as the density gradient, Laplacian, kinetic-energy density, and Hirshfeld atomic partitioning weights. The $f_n$ damping functions depend on the interatomic distance, $R_{ij}$, and will be discussed in detail in the remainder of this section.

Conventionally, XDM uses the Becke--Johnson (BJ) damping function,\cite{johnson2006post} which is also used in the D3(BJ) and D4 dispersion methods of Grimme and co-workers.
%\kyle{While probably fine. there is slight repeating from two paragraphs up.: ``The error was traced to the Becke–Johnson (BJ) damping function 19 used in XDM (as well as in the D3(BJ) and D4 dispersion models)''}
This damping function is given by
\begin{equation}
f_n^\text{BJ}(R_{ij}) = \frac{R^n_{ij}}{R^n_{ij} + R^n_{\text{vdW},ij}} \,,
\end{equation}
where $R_{\text{vdW},ij}$ is the sum of approximate van der Waals radii of atoms $i$ and $j$. It is determined as
\begin{equation}
R_{\text{vdW},ij} = a_1 R_{\text{c},ij} + a_2 \,,
\end{equation}
where $a_1$ and $a_2$ are empirical parameters that are not element-dependent but are fitted for use with a particular combination of density functional and basis set. $R_{\text{c},ij}$ is a ``critical'' interatomic distance at which successive terms in the perturbation theory expansion of the dispersion energy become equal. If the dispersion energy only includes the $C_6$ and $C_8$ terms, then 
\begin{equation}
R_{\text{c},ij} = \sqrt{\frac{C_{8,ij}}{C_{6,ij}}} \,.
\end{equation}
However, if the $C_{10}$ term is also included in the dispersion energy, two other possible definitions for $R_{\text{c},ij}$ arise:
\begin{equation}
R_{\text{c},ij} = \left\{ \begin{array}{l} \sqrt{\frac{C_{10,ij}}{C_{8,ij}}} \\[8pt] \sqrt[4]{\frac{C_{10,ij}}{C_{6,ij}}} \end{array} \right. \,.
\end{equation}
In XDM, the value of $R_{\text{c},ij}$ is taken to be the average of these three results: 
%\kyle{Suggest showing the equation since the paper is about comparing the damping functions
\begin{equation}
R_{\text{c},ij} = \frac{1}{3} \left[ \left( \frac{C_{8,ij}}{C_{6,ij}} \right)^{1/2} +  \left( \frac{C_{10,ij}}{C_{6,ij}} \right)^{1/4} +  \left( \frac{C_{10,ij}}{C_{8,ij}} \right)^{1/2}\right] \,.
\end{equation} 
%}

Becke recently proposed an alternative damping function for use with XDM that, unlike BJ damping, involves only one empirical fit parameter.\cite{becke2024remarkably} In this work, it will be referred to as Z damping, due to the dependence on the atomic number. The Z-damping function is 
\begin{equation}\label{eq:zdamp}
f_n^\text{Z} (R_{ij}) = \frac{R^n_{ij}}{R^n_{ij} + z_\text{damp}
\frac{C_{n,ij}}{Z_i+Z_j}} \,,
\end{equation}
where $Z_i$ and $Z_j$ are the atomic numbers of atoms $i$ and $j$, respectively. This definition was chosen because the resulting contribution to the correlation energy in the united-atom limit would be
\begin{equation}
\lim_{R_{ij}\rightarrow0} \left( \frac{C_{n,ij}}{R_{ij}^n + z_\text{damp} \frac{C_{n,ij}}{Z_i+Z_j}} \right) =  \frac{Z_i + Z_j}{z_\text{damp}} \,,
\end{equation}
and atomic correlation energies are roughly proportional to atomic number.\cite{burke2016locality} Similar to BJ damping, the single empirical parameter, $z_\text{damp}$, is atom-independent and fitted for use with a particular density functional and basis set. A typical value of $z_\text{damp}$ is around $10^{5}~\text{Ha}^{-1}$.

\begin{figure}
\includegraphics[width=\columnwidth]{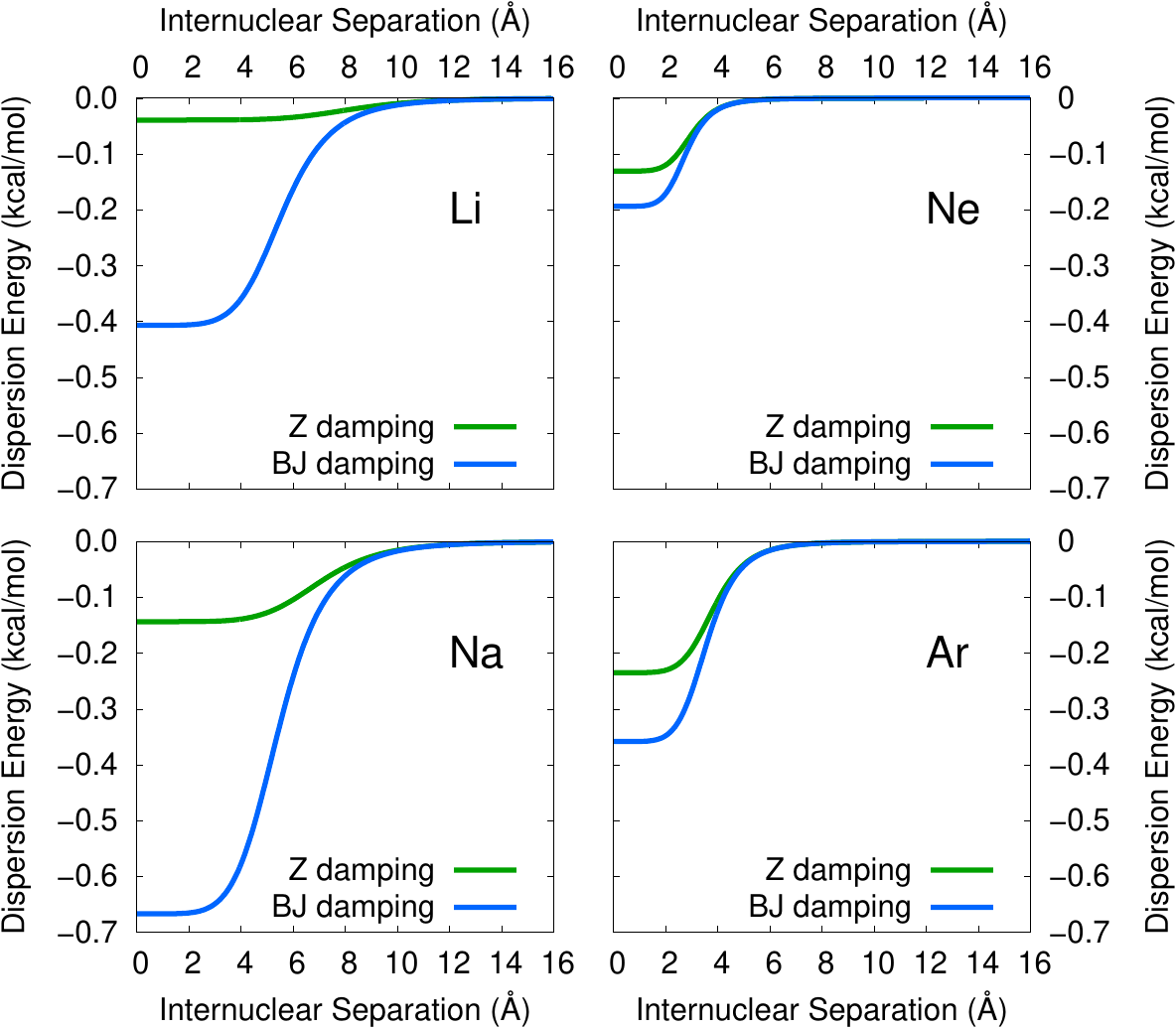}
\caption{Comparison of BJ- and Z-damping functions. The plots use XDM data for the free atoms only, computed with the B86bPBE functional and \texttt{tight} basis settings using FHI-aims.}
\label{fig:damp}
\end{figure}

To illustrate the differences in damping functions, BJ  and Z damping are compared for homonuclear interactions between Li, Na, Ne, and Ar atoms in Figure~\ref{fig:damp}. The Li and Na calculations are spin-polarized, with a net spin of 1 electron. For simplicity, the curves use only data for the free atoms, which omit changes in dispersion coefficients with internuclear separation that would be observed in the dimer systems due to varying electron densities. The results in Figure~\ref{fig:damp} show that Z damping consistently reduces the magnitude of the dispersion energy compared to BJ damping. However, this effect is fairly minor for Ne and Ar, while there is a very large increase in damping strength for Li and Na. This allows correction of the overbinding seen with BJ damping for the Li$_8$ and Na$_8$ clusters, while preserving high accuracy for main-group elements. With BJ damping, the magnitudes of the dispersion energies in the united-atom limit follow the trend Na$>$Li$>$Ar$>$Ne, but this changes to Ar$>$Na$\approx$Ne$>$Li for Z damping. The latter appears more physical because, in the united-atom limit, the dispersion energy would become a correlation energy and should increase with the number of electrons and, hence, atomic number.\cite{burke2016locality}

\section{Data Sets}

To evaluate the performance of XDM with both BJ and Z damping, a comprehensive list of benchmarks has been selected for testing. These are categorised into three groups: those used to optimize parameters for damping functions, finite-molecule benchmarks, and molecular-crystal benchmarks. The benchmark content, geometry sources, and reference data quality are summarised below.

\subsection{Damping-Parameter Fit Set}\label{ss:c6}

\textbf{KB49}:
Binding energies of 49 molecular dimers with reference values from basis-set extrapolated CCSD(T) calculations.\cite{kannemann2010van} Dimer geometries are available from the \texttt{refdata} GitHub repository.\cite{otero2015refdata} The BJ-damping parameters, $a_1$ and $a_2$ (in \AA), as well as the Z-damping parameter, $z_\text{damp}$, were fitted separately for each combination of DFA and basis set. Optimal parameters were determined by minimising the root-mean-square percent error (RMSPE) for the KB49 set. 
%\kyle{Removing: "and may be found in the ESI." to avoid repetition with the first paragraph of Computational Methods}

\subsection{Molecular Benchmarks}

\textbf{GMTKN55}: A collection of 55 individual benchmarks spanning the thermochemistry of small and large molecules, reaction barriers, and both intramolecular and intermolecular non-covalent interactions. 
Ref.~\citenum{goerigk2017look} provides detailed information regarding the individual benchmarks. Geometries for FHI-aims may be obtained from the \texttt{gmtkn55-fhiaims} GitHub repository.\cite{johnson2024gmtkn55} 

Due to the wide range of energy scales across the component benchmarks within GMTKN55, the overall error is reported as a weighted mean absolute deviation (\mbox{WTMAD}). While several weighted error definitions have been proposed, this work focuses on our recent WTMAD-4, introduced in Ref.~\citenum{bryenton2026wtmad4}. WTMAD-4 was developed to address the disproportionate weighting observed in earlier \mbox{WTMAD} schemes, which arose from coupling each benchmark's weight to its mean reference energy, $| \overline{\Delta E} |$. Analysis of mean errors for a test set of 115 dispersion-corrected functionals (ranging from GGAs to double hybrids) showed that the canonical WTMAD-2 scheme weighed some benchmarks more than $200\times$ greater than others, with the top 3 benchmarks contributing as much to the total WTMAD-2 as the bottom 36. 

In WTMAD-4, each subset is assigned a weight, denoted $w_{i}$, based on the mean error ($\overline{\text{MAD}}_i$) obtained from a set of 10 representative minimally empirical hybrid functionals:
\begin{equation}
w_i = \frac{100}{N_{\text{bench}}} \left( \frac{3.5}{\overline{\text{MAD}}_{i}^{\text{10-DFA}}}\right) \,.
\end{equation}
The \mbox{WTMAD-4} is then defined as
\begin{equation} \label{eq:wtmad4}
\text{WTMAD-4} = \frac{1}{N_\text{bench}} \sum_{i=1}^{N_\text{bench}} w_{i} \cdot \text{MAD}_{i} \,.
\end{equation}
This weighted error definition substantially improves the weighting balance, such that benchmarks are typically weighted within a factor of $\sim$3 of one another, and the top 3 benchmarks now contribute comparably to the bottom 6. See the ESI for the specific values of the weights, and Ref.~\citenum{bryenton2026wtmad4} for more information regarding the formulation of the WTMAD-4 and application to data compiled from Refs.~\citenum{becke2024remarkably, goerigk2017look, santra2021exploring} and \citenum{wittmann2023dispersion}.

In this work, we also report the number, $N_{r>h}$, of benchmarks that have a ratio of $\nicefrac{\text{MAD}_i}{\,\overline{\text{MAD}}_{i}^{\text{10-DFA}}}>h$, where the mean MAD is the value obtained from the 10 reference DFAs used in the definition of the WTMAD-4 weights. Similarly, $N_{d>h}$ is the number of benchmarks that have the difference $\text{MAD}_i- \overline{\text{MAD}}_{i}^{\text{10-DFA}}> h$ in units of kcal/mol. We would ideally want $N_{r\leq1} = N_{d\leq0}$ to be 55, meaning equal or better performance than the average of the chosen 10 functionals for all of the benchmarks, but barring that, we seek to avoid any extreme outliers in terms of both absolute and percent errors by minimising both $N_{d>2}$ and $N_{r>2}$. These criteria, qualitatively, identify cases where the error more than doubles relative to the 10-DFA mean, or exceeds it by more than 2 kcal/mol.

\subsection{Solid-State Benchmarks}

\textbf{X23}:  %\cite{reilly2013understanding, otero2012benchmark, dolgonos2019revised}}
Lattice energies of 23 molecular crystals,\cite{reilly2013understanding,otero2012benchmark} using updated ``X23b'' reference energies.\cite{dolgonos2019revised}  Geometries are available from the \texttt{refdata} repository.\cite{otero2015refdata} Unlike the previous benchmarks, X23 requires geometry optimisations with each functional and basis combination considered. 

\textbf{HalCrys4}:  %\cite{otero2019dispersion, dean1999lange}}
Lattice energies of four halogen crystals---\ch{Cl2}, \ch{Br2}, \ch{I2}, and \ch{ICl}.\cite{otero2019dispersion} The lattice energies are compared to back-corrected experimental results from Ref.~\citenum{dean1999lange}. As with the X23, geometries are optimised for each reported functional and basis set. Geometries are available from the \texttt{refdata} repository.\cite{otero2015refdata}

\textbf{ICE13}:  
Absolute lattice energies of ice polymorphs \cite{brandenburg2015benchmarking}
(Abs), along with their relative energy differences (Rel) using diffusion Monte Carlo (DMC) reference data.\cite{della2022dmc} ICE13 requires geometry optimisations for all systems except the isolated water molecule, which uses a fixed geometry. Geometries are available from the \texttt{refdata} repository.\cite{otero2015refdata}

\section{Computational Methods} \label{ss:compmethods}

\begin{table*}[ht]
\caption{Optimum XDM(BJ) and XDM(Z) damping parameters for use with the \texttt{lightdenser} and \texttt{tight} basis sets.}
\label{t:damping}
\centering
\begin{tabular}{l|ccc|ccc} \hline
Functional & \multicolumn{3}{c|}{lightdenser} & \multicolumn{3}{c}{tight} \\ 
 & $a_1$ & $a_2$ (\AA) & $z_\text{damp}$ & $a_1$ & $a_2$ (\AA) & $z_\text{damp}$  \\ \hline
 PBE                 & 0.3275 & 2.9627 & 200770 & 0.5124 & 2.2588 & 162373 \\
 B86bPBE             & 0.6881 & 1.5789 & 116996 & 0.9004 & 0.7808 & ~96089 \\ 
 revPBE              & 0.9255 & 0.3649 & ~39880 & 0.8992 & 0.2849 & ~32842 \\ \hline
 PBE0                & 0.1775 & 3.5217 & 238489 & 0.4713 & 2.3855 & 162110 \\
 B86bPBE0            & 0.4545 & 2.4309 & 153336 & 0.7284 & 1.3781 & 108291 \\
 revPBE0             & 0.5358 & 1.7557 & ~65559 & 0.7495 & 0.9199 & ~48549 \\
 B3LYP               & 0.5816 & 1.7060 & ~78928 & 0.6791 & 1.2394 & ~59992 \\ \hline
 PBE50               & 0.$^e$ & 4.3052 & 353058 & 0.4233 & 2.5711 & 173943 \\
 B86bPBE50           & 0.0330 & 3.9929 & 250118 & 0.5908 & 1.9047 & 131706 \\
 revPBE50            & 0.$^e$ & 3.7113 & 132297 & 0.5157 & 1.8595 & ~76298 \\                 
 BHLYP               & 0.$^e$ & 3.8799 & 173138 & 0.2877 & 2.7329 & 106257 \\ \hline
 HSE06               & 0.1579 & 3.6101 & 250809 & 0.4523 & 2.4809 & 173227 \\
 LC-$\omega$PBE$^a$  & 0.5012 & 2.2201 & 134418 & 0.7547 & 1.2843 & 105799 \\
 LC-$\omega$PBE$^b$  & 0.3553 & 2.7882 & 169597 & 0.9496 & 0.7045 & 110361 \\
 LC-$\omega$hPBE$^c$ & 0.5271 & 2.2492 & 164070 & 0.6849 & 1.5553 & 117529 \\
 LC-$\omega$hPBE$^d$ & 0.4094 & 2.7666 & 218061 & 0.8571 & 1.0526 & 124311 \\ \hline
\end{tabular}\\
$^a$ $\omega=0.2$; $^b$ $\omega=0.4$; $^c$ $\omega=0.2$ and $a_X=0.2$; $^d$ $\omega=0.4$ and $a_X=0.2$;\\
$^e$Set to zero to prevent unphysical, negative values.
\end{table*}

All calculations were performed using versions 250425 or 250711 of FHI-aims.\cite{blum2009ab, ren2012resolution, levchenko2015hybrid, kokott2024efficient, yu2018elsi, havu2009efficient, ihrig2015accurate, price2023xdm} 
These versions of FHI-aims support integer \texttt{occupation\_type} for spin-polarized systems, which is required for obtaining appropriate densities and energies for lone atoms (and certain dimers) used in some benchmarks, such as the W4-11 atomization energies subset. We note that later versions of FHI-aims have removed this capability as it's an area of active development. The scalar zeroth-order regular approximation (ZORA) relativistic correction\cite{van1994relativistic} was used throughout. As noted above, the BJ- and Z-damping coefficients were determined for each functional and basis combination by least-squares fitting to minimise the RMSPE for the KB49 benchmark set of intermolecular binding energies. Parameters for the XDM(BJ) and XDM(Z) dispersion corrections optimised for combinations of 16 density functionals and two basis sets are shown in Table~\ref{t:damping}, 
and all damping function parameterisations at the time of writing are included in the ESI. A regularly updated list of all XDM BJ- and Z-damping parameters for various functional--basis combinations is kept in the \texttt{refdata} GitHub repository.\cite{otero2015refdata} For a version of FHI-aims that automatically sets the XDM damping parameters for all functionals considered here, the interested reader is directed to versions 260110 onwards.

At the GGA level of theory, we considered the PBE,\cite{becke1986density, perdew1996generalized, perdew1997erratum} revPBE,\cite{zhang1993comment} and B86bPBE\cite{becke1986large} functionals. At the global hybrid level, we selected several GGA-based hybrids including B3LYP,\cite{becke1988density, beck1993density, lee1988development, stephens1994ab, vosko1980accurate} popular for molecular thermochemistry; PBE0,\cite{adamo1999toward} popular in solid-state chemistry; revPBE0,\cite{zhang1993comment,adamo1999toward} popular for studies of water; and our previously recommended B86bPBE0.\cite{price2023xdm} We also used the analogues of these functionals with 50\% exact exchange (BHLYP\cite{becke1993new}, PBE50, revPBE50, and B86bPBE50), which should exhibit reduced delocalisation error.\cite{bryenton2023delocalization} 
Finally, we considered the range-separated GGA-based hybrid, HSE06,\cite{krukau2006influence} and four parameterisations of the \mbox{LC-$\omega$(h)PBE} functional.\cite{vydrov2006assessment, vydrov2007tests}

It is notable that the TS,\cite{tkatchenko2009accurate} MBD@rsSCS,\cite{ambrosetti2014long} and MBD-NL\cite{hermann2020density} dispersion corrections, also available in FHI-aims, have not been tested for the GMTKN55 benchmark. As a result, calculations were performed using each of these three dispersion corrections, paired with only the PBE and PBE0 functionals due to the limited availability of damping parameters. Additionally, while D3(BJ) has been widely applied in the literature,\cite{goerigk2017look} the MAE data is not available for its pairing with revPBE0 specifically, which is found to be one of the top-performing hybrids.\cite{santra2019minimally} Thus, D3(BJ) calculations were performed for the PBE, PBE0, revPBE, and revPBE0 functionals using FHI-aims. This allows comparison between FHI-aims (this work) and Gaussian-basis (Ref.~\citenum{goerigk2017look}) results for the other three functionals.

For GMTKN55, all FHI-aims calculations used the \texttt{tight} basis, except for subsets containing anions. HB21, BH76, BH76RC, and G21EA used \texttt{tier2\_aug2} for all atoms; IL16 used \texttt{tier2\_aug2} for all O, F, S, and Cl atoms; and WATER27 used \texttt{tier2\_aug2} for O atoms only for reactions involving anions, as this basis caused linear dependencies in the SCF for some of the larger, neutral water clusters. In all cases, the damping parameters were kept at the same values optimised for the \texttt{tight} basis settings as these are already sufficiently converged as to approach the basis-set limit. 

Turning to the solid-state, only the three GGA and six global-hybrid functionals were considered (B3LYP and BHLYP were omitted as the asymptotic constraint used in the construction of the B88 exchange functional\cite{becke1988density} is not relevant for solid-state systems). The GGA calculations used both the \texttt{tight} and \texttt{lightdenser} basis settings as the latter is our recommended basis for most solid-state calculations (particularly geometry optimisations), although there will be some residual basis-set incompleteness error. For the hybrid functionals, only \texttt{lightdenser} calculations were performed as calculations with the \texttt{tight} basis require prohibitive amounts of memory. Hybrid results with the \texttt{tight} basis were approximated using an additive basis set correction evaluated at the converged GGA/\texttt{lightdenser} geometries:\cite{hoja2018first,price2023xdm}
\begin{align}\label{eq:bsc}
E(\text{hybrid/\texttt{tight}}) & \approx E(\text{hybrid/\texttt{lightdenser}}) \notag\\
& + E(\text{GGA/\texttt{tight}}) \notag\\
&- E(\text{GGA/\texttt{lightdenser}}) \,.
\end{align}

As previously mentioned, this work employs the new \texttt{lightdenser} basis, which builds on the \texttt{lightdense} basis introduced in Ref.~\citenum{price2023xdm} and is now packaged in the species defaults of the FHI-aims code. \texttt{lightdense} uses \texttt{light} basis functions and increases the integration grids to those of the \texttt{tight} basis defaults, removing instabilities that resulted in artificial minima in the potential energy surface and could sometimes prevent convergence of geometry optimisations to their true minima. The \texttt{lightdenser} basis builds on this by also increasing the highest angular momentum component present in the multipole expansion of the Hartree potential (\texttt{l\_hartree}) to 8, resolving a small force--energy inconsistency that could, on rare occasions, also prevent convergence of geometry optimisations. This \texttt{lightdenser} basis offers increased stability while incurring only slightly more computation time and a negligible increase in memory requirements compared to its \texttt{light} counterpart. We recommend this \texttt{lightdenser} basis, particularly for solid-state applications or in cases where the \texttt{tight} basis is prohibitively large.

Lastly, we highlight the computational efficiency of the XDM-based post-SCF dispersion corrections. These corrections account for only a small fraction of the total CPU time. Typically, using XDM (with either BJ or Z damping) requires less CPU time than even 10\% of a single SCF step.

\section{Results and Discussion}

\subsection{Molecular Benchmarks}

The focus of this section is the GMTKN55 set, comprised of 55 diverse molecular benchmarks. 
Table~\ref{tab:shaded} shows a detailed comparison of the performance of XDM(BJ) versus XDM(Z) for each of the component benchmarks using three selected DFAs.
Full statistics for each benchmark with all functionals and dispersion corrections, as well as the WTMAD-$N$ values for each category, are provided in the ESI.

\begin{table}[ht!]
\caption{Comparison of MADs (in kcal/mol) for the individual GMTKN55 benchmarks using the selected functionals with either XDM(BJ) or XDM(Z). Also shown for comparison are the $\overline{\text{MAD}}_{i}^{\text{10-DFA}}$ values from the 10 representative DFAs used in the definition of the WTMAD-4; these mean values are used to quantify outliers. Entries are shaded according to their difference from the $\overline{\text{MAD}}_{i}^{\text{10-DFA}}$ values. 
}\label{tab:shaded}
\centering
\includegraphics[width=\columnwidth]{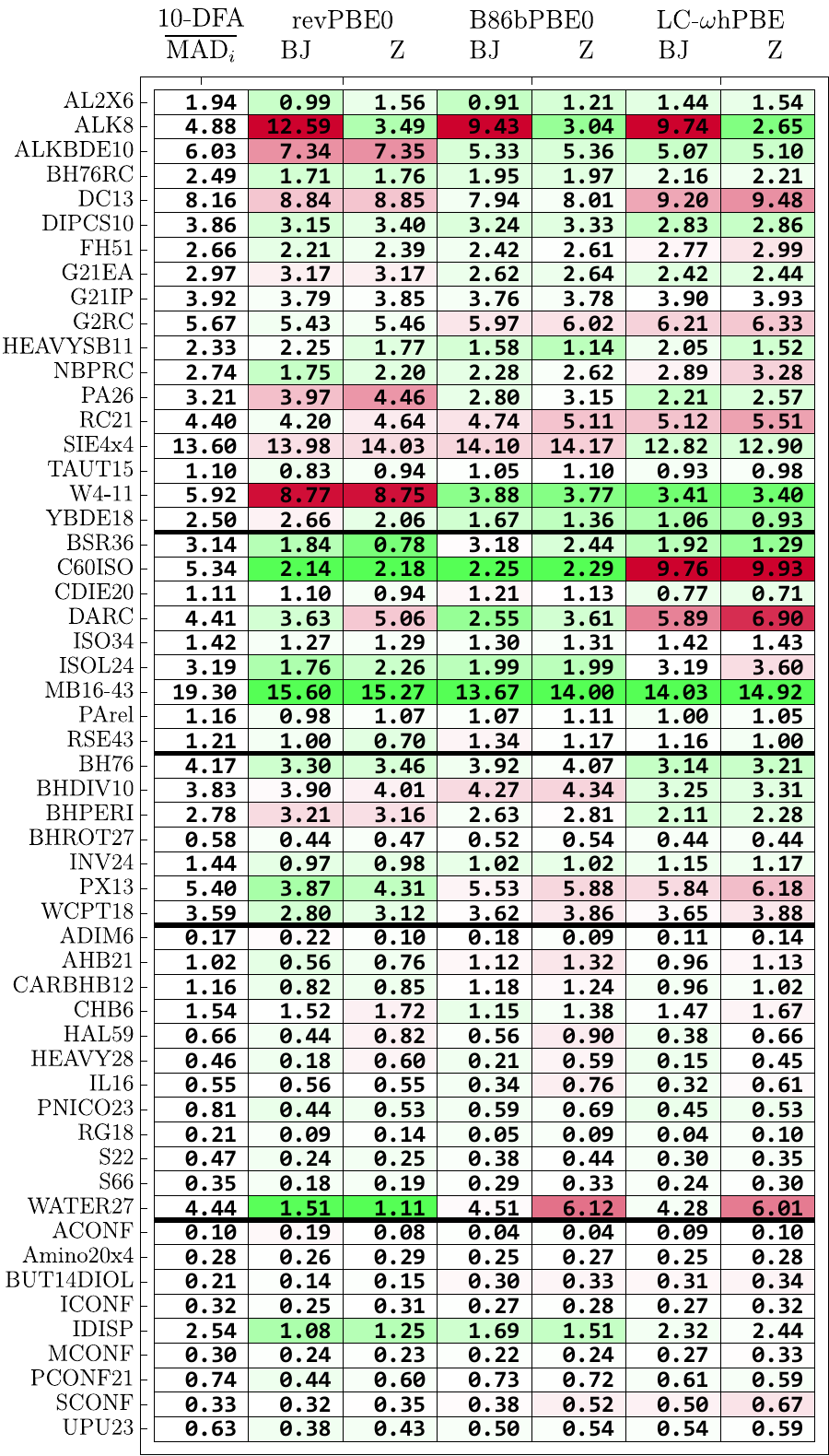}
\end{table}

\begin{table*}[t]
\caption{\mbox{WTMAD-4} results, numbers of outliers, and maximum outliers, for the GMTKN55 benchmark for selected functionals and dispersion corrections. 
$N_{r>h}$ is the number of benchmarks that have a ratio of $\nicefrac{\text{MAD}_i}{\,\overline{\text{MAD}}_{i}^{\text{10-DFA}}}>h$, where the mean MAD is the value obtained from the 10 reference DFAs used in the definition of the WTMAD-4 weights. Similarly, $N_{d>h}$ is the number of benchmarks that have the difference $\text{MAD}_i- \overline{\text{MAD}}_{i}^{\text{10-DFA}}> h$ kcal/mol.}
\label{tab:gmtkn55}
\centering
{\footnotesize
\renewcommand{\tabcolsep}{2pt}
\begin{tabular}{l|ccrrrlrrrl|ccrrrlrrrl}\hline
Functional           & \multicolumn{2}{c}{WTMAD-$N$} & \multicolumn{8}{c|}{XDM(BJ)} & \multicolumn{2}{c}{WTMAD-$N$} & \multicolumn{8}{c}{XDM(Z)} \\ 
                     & 2    & 4    & $N_{r\leq1}$ & $N_{r>2}$ & Max $r$ & Set & $N_{d>2}$ & $N_{d>5}$ & MAX $d$ & Set
                     & 2    & 4    & $N_{r\leq1}$ & $N_{r>2}$ & Max $r$ & Set & $N_{d>2}$ & $N_{d>5}$ & MAX $d$ & Set \\ \hline
PBE                  &10.39 & 9.20 & 11 & 12 & 2.79 & ALK8      & 13 & 8 & 9.86  & SIE4x4  &10.77 & 9.27 & 11 & 13 & 2.62 & W4-11     & 12 & 7 & 9.96  & SIE4x4  \\
B86bPBE              & 9.37 & 8.48 & 16 & 11 & 2.94 & SCONF     & 11 & 4 & 9.80  & SIE4x4  & 9.99 & 8.80 & 14 & 13 & 2.99 & SCONF     & 11 & 4 & 9.82  & SIE4x4  \\ 
revPBE               & 8.40 & 8.45 & 16 & 7  & 3.76 & ACONF     & 11 & 4 & 9.79  & SIE4x4  & 8.69 & 7.91 & 19 & 7  & 2.51 & SCONF     & 8  & 3 & 9.65  & SIE4x4 \\ \hline
PBE0                 & 6.55 & 6.21 & 30 & 1  & 3.19 & ALK8      & 1  & 1 & 10.68 & ALK8    & 6.96 & 6.19 & 28 & 0  & 1.64 & WATER27   & 1  & 0 & 2.82  & WATER27 \\
B86bPBE0             & 5.77 & 5.53 & 39 & 0  & 1.93 & ALK8      & 1  & 0 & 4.55  & ALK8    & 6.43 & 5.85 & 39 & 0  & 1.55 & SCONF     & 0  & 0 & 1.68  & WATER27 \\
revPBE0              & 5.06 & 5.44 & 42 & 1  & 2.58 & ALK8      & 2  & 1 & 7.71  & ALK8    & 5.71 & 5.43 & 40 & 0  & 1.48 & W4-11     & 1  & 0 & 2.83  & W4-11   \\
B3LYP                & 6.26 & 6.27 & 36 & 2  & 2.44 & BUT14DIOL & 5  & 2 & 6.00  & MB16-43 & 6.87 & 6.53 & 31 & 1  & 2.46 & BUT14DIOL & 3  & 1 & 6.88  & MB16-43 \\ \hline
PBE50                & 6.05 & 7.44 & 20 & 5  & 3.60 & ALK8      & 7  & 5 & 12.67 & ALK8    & 6.43 & 7.31 & 24 & 4  & 2.46 & W4-11     & 6  & 4 & 8.63  & W4-11 \\
B86bPBE50            & 5.66 & 6.98 & 25 & 5  & 2.71 & W4-11     & 7  & 4 & 10.15 & W4-11   & 6.19 & 7.08 & 23 & 4  & 2.70 & W4-11     & 6  & 4 & 10.05 & W4-11 \\
revPBE50             & 5.40 & 6.97 & 27 & 4  & 4.42 & ALK8      & 8  & 4 & 16.70 & ALK8    & 5.84 & 6.73 & 28 & 4  & 3.37 & W4-11     & 6  & 4 & 14.02 & W4-11   \\
BHLYP                & 5.71 & 7.59 & 27 & 6  & 6.56 & ALK8      & 9  & 5 & 27.14 & ALK8    & 6.07 & 7.46 & 26 & 5  & 3.34 & W4-11     & 9  & 4 & 18.82 & MB16-43 \\ \hline
HSE06                & 6.85 & 6.41 & 31 & 1  & 3.23 & ALK8      & 1  & 1 & 10.90 & ALK8    & 7.23 & 6.37 & 31 & 0  & 1.87 & BUT14DIOL & 1  & 0 & 3.75  & WATER27 \\
LC-$\omega$PBE$^a$   & 6.92 & 6.56 & 35 & 2  & 2.84 & SCONF     & 5  & 2 & 5.77  & SIE4x4  & 7.60 & 7.10 & 29 & 2  & 3.22 & SCONF     & 6  & 2 & 5.84  & SIE4x4  \\ 
LC-$\omega$PBE$^b$   & 5.31 & 6.39 & 32 & 2  & 3.17 & C60ISO    & 4  & 2 & 11.61 & C60ISO  & 5.65 & 6.57 & 31 & 3  & 3.16 & C60ISO    & 4  & 2 & 11.52 & C60ISO \\ 
LC-$\omega$hPBE$^c$  & 5.32 & 5.65 & 42 & 0  & 2.00 & ALK8      & 2  & 0 & 4.86  & ALK8    & 5.99 & 6.09 & 34 & 0  & 2.00 & SCONF     & 2  & 0 & 4.59  & C60ISO  \\ 
LC-$\omega$hPBE$^d$  & 5.70 & 7.10 & 23 & 4  & 3.73 & C60ISO    & 7  & 3 & 14.57 & C60ISO  & 6.04 & 7.23 & 25 & 4  & 3.73 & C60ISO    & 8  & 3 & 14.56 & C60ISO \\ \hline
\end{tabular}\\
}
$^a$ $\omega=0.2$; $^b$ $\omega=0.4$; $^c$ $\omega=0.2$ and $a_X=0.2$; $^d$ $\omega=0.4$ and $a_X=0.2$.
\end{table*}

From the results in Table~\ref{tab:shaded},
Z damping shows clear improvements for ALK8 (dissociation and other reactions of alkaline compounds), HEAVYSB11 (dissociation energies of heavy-element compounds), YBDE18 (bond-dissociation energies of ylides), and BSR36 (bond-separation reactions of saturated hydrocarbons). Conversely, BJ damping performs better for DARC (Diels-Alder reaction energies), NBPRC (oligomerisations, H$_2$ fragmentations, and H$_2$ activation reactions involving NH$_3$/BH$_3$ or PH$_3$/BH$_3$ systems), PA26 (adiabatic proton affinities), and RC21 (fragmentations and rearrangements in radical cations). For most other subsets, there is little to choose between the two damping schemes. 

Comparing the three selected functionals, B86bPBE0 consistently achieves the minimum error on the MB16-43  (mindless benchmarking) subset, with MAEs of 13.7 and 14.0 kcal/mol; for comparison, it has been noted that ``MADs for MB16-43 usually exceed 15 kcal/mol for most dispersion-corrected hybrid DFAs.''\cite{goerigk2017look} In terms of outliers, LC-$\omega$hPBE is conspicuously poor for C60ISO (relative energies of C$_{60}$ isomers), while revPBE0 gives large errors for W4-11 (total atomisation energies), ALKBDE10 (dissociation energies in group-1 and -2 diatomics), and PA26. It makes sense that revPBE exchange is poor for atomisation energies since, unlike most exchange functionals, it was not fit to atomic exchange energies.\cite{zhang1993comment} However, revPBE0 still yields the lowest WTMAD-4 of all XDM-corrected functionals due to its excellent performance for WATER27 (binding energies of water clusters). Replacing BJ damping with Z damping slightly improves the performance for WATER27 using revPBE0, but significantly increases the overbinding seen with B86bPBE0 and LC-$\omega$hPBE.

The overall WTMAD-4 values obtained for the GMTKN55 using all XDM-corrected functionals considered are summarised in Table~\ref{tab:gmtkn55}. The table also shows the distribution of MAD$_i$ values, as indicated by selected $N_{r>h}$ and $N_{d>h}$ metrics.
As expected, the GGA functionals show larger errors than the hybrid and range-separated hybrid functionals, with maximum differences from the representative means seen for the SIE4x4 set (systems with large self-interaction errors). In terms of the WTMAD-4, B86b and revPBE exchange generally outperform PBE exchange, which reinforces our previous conclusion as to the importance of using a dispersionless DFA in combination with post-SCF dispersion corrections.\cite{price2021requirements} Despite reduced self-interaction error, the 50\% hybrid functionals offer overall poorer performance than the 20-25\% hybrids, with the W4-11 set of atomisation energies consistently being a large outlier. Finally, the WTMAD-4's for the range-separated hybrids are quite sensitive to the choice of range-separation parameter and neglect or inclusion of short-range exact-exchange mixing. The C60ISO set is a large outlier for all the RS functionals with full long-range exact exchange.

The best performing XDM-based method overall is revPBE0-XDM(Z), which gives the lowest WTMAD-4 (5.43) and, importantly, no large outliers with errors greater than 2$\times$ the mean MAD obtained with our 10 reference DFAs
($N_{r>2}=0$), although W4‐11   remains a significant outlier in terms of absolute error (2.83~kcal/mol above the mean MAD). While revPBE0-XDM(BJ) has a similar WTMAD-4, ALK8 remains a very large outlier in terms of both relative and absolute errors. B86bPBE0-XDM(Z) is the second best-performing combination overall, with a higher WTMAD-4 (5.85), but having $N_{r>2}=0$ and also $N_{d>2}=0$; the largest absolute error occurs for WATER27, where the MAD is 1.68~kcal/mol above the mean MAD. Thus, the choice of revPBE0-XDM(Z) versus B86bPBE0-XDM(Z) may come down to whether the user prefers greater accuracy for atomisation energies, or for water clusters, which will depend on whether they are modelling covalent or non-covalent chemistry.

\subsection{Comparison with Literature Functionals}

Table~\ref{tab:rankings} shows the top ranked DFAs available for each ``rung'' of Perdew's ladder\cite{perdew2001jacob} according to lowest \mbox{WTMAD-4} values, combining the current data with that from Ref.~\citenum{bryenton2026wtmad4} and new additional data from Refs.~\citenum{najibi2018nonlocal, najibi2020dft, luise2025accurate, kirkpatrick2021pushing}. It has been previously demonstrated that the numerical atomic orbital (NAO) basis sets used in FHI-aims give energies in good agreement with basis-set limit results using Gaussian-type orbitals (GTOs).\cite{abbott2025roadmap,elephant} Indeed, the extremely similar metrics for revPBE-D3(BJ) 
(viz.\ WTMAD-4 values of 7.67 vs.\ 7.73, maximum ratio outliers of 2.22 vs.\ 2.26, and maximum difference outliers of 9.70 vs.\ 9.83 kcal/mol) shown in the ESI confirm that NAO and GTO results are directly comparable. However, it is notable that the \texttt{tight} basis set in FHI-aims includes fewer functions than the typical \texttt{def2-QZVPP(D)} basis used for the GMTKN55 benchmark, yet delivers nearly the same results.

From the data in Table~\ref{tab:rankings}, XDM-corrected functionals show consistently strong performance when paired with GGAs or GGA-based global hybrids. In terms of \mbox{WTMAD-4} values, the D3(BJ), XDM(BJ), and XDM(Z) dispersion corrections all perform similarly; it is principally for metal-containing benchmarks, such as ALK8, where there are notable differences that manifest in the distribution of outliers. Each of these three dispersion corrections, paired with revPBE or revPBE0, is a top-performing functional within its class. 
While fitting no parameters and using only GGA ingredients in the base DFAs, revPBE0 with D3(BJ), XDM(Z), and XDM(BJ) ranks 5th through 7th in terms of lowest \mbox{WTMAD-4} values for global hybrids. Although slightly lower \mbox{WTMAD-4} values are obtained with PW6B95-D3(BJ)/V,\cite{pw6b95} M05-2X-D3(0),\cite{m052x} and M06-2X-D3(0),\cite{m062x} these base functionals involve 6, 23, and 33 empirical fit parameters, respectively. They also have much more complicated functional forms, relying on high-order power-series expansions and/or meta-GGA ingredients that, in turn, give rise to numerical instabilities.\cite{cpl,oscillations,wheeler,headgordon} Additionally, M05-2X-D3(0) and M06-2X-D3(0) both give large difference outliers (for the MB16-43 and HEAVYSB11 sets, respectively), which is likely indicative of overfitting.

\begin{table*}[t!]
\caption{The top 6 GGAs, 6 meta-GGAs, 12 global hybrids, 12 range-separated hybrids, 6 double-hybrid functionals, and 5 machine-learned functionals, sorted according to WTMAD-4. Also shown are the numbers of outliers and the GKMTKN55 subset responsible for the maximum outlier in terms of both ratio and difference errors.  WTMAD-2 values are also listed for comparison; as discussed in Ref.~\citenum{bryenton2026wtmad4}, the WTMAD-2 literature values may have a margin of error of $\sim$1-2\% due to ambiguity of specific reference values used and unavailability of complete benchmark results from literature. If the same functional-DC combination was listed in multiple sources, only the lowest WTMAD-4 value and corresponding outlier statistics are quoted here.}\label{tab:rankings}
\centering
{\small
\begin{tabular}{l|cc|rrrlrrrl}\hline
& \multicolumn{2}{c|}{WTMAD-$N$} \\
Functional                                              & 2 & 4 & $N_{r\leq1}$ & $N_{r>2}$ & MAX $r$ & Set & $N_{d>2}$ & $N_{d>5}$ & MAX $d$ & Set       \\
\hline
revPBE-D3(BJ)$^a$ \cite{zhang1993comment}                                         & 8.24 & 7.67 & 16 & 2 & 2.22 & BHPERI    &  9 & 2 &  9.70 & SIE4x4    \\
%revPBE-D3(BJ)$^{b,c}$ \cite{zhang1993comment}                                    & 8.29 & 7.73 & 16 & 3 & 2.26 & BHPERI    &  8 & 2 &  9.83 & SIE4x4    \\
revPBE-XDM(Z)$^a$ \cite{zhang1993comment}                                         & 8.69 & 7.91 & 19 & 7 & 2.51 & SCONF     &  8 & 3 &  9.65 & SIE4x4    \\
revPBE-V$^c$ \cite{zhang1993comment}                                              & 8.50 & 7.94 & 23 & 8 & 2.88 & DIPCS10   & 11 & 6 &  9.36 & MB16-43   \\
OLYP-D3(BJ)$^b$ \cite{optx,lee1988development}                                    & 8.76 & 8.16 & 18 & 6 & 2.51 & BHPERI    & 13 & 2 & 11.95 & SIE4x4    \\
B97-D3(BJ)$^b$ \cite{grimme2006semiempirical}                                     & 8.49 & 8.32 & 19 & 6 & 2.57 & AL2X6     & 12 & 4 & 16.80 & MB16-43   \\
revPBE-XDM(BJ)$^a$ \cite{zhang1993comment}                                        & 8.40 & 8.45 & 16 & 7 & 3.76 & ACONF     & 11 & 4 &  9.79 & SIE4x4    \\
\hline
B97M-V$^{c,e}$ \cite{b97mv}                                                       & 5.46 & 5.24 & 39 & 0 & 1.86 & MB16-43   &  2 & 1 & 16.67 & MB16-43   \\
B97M-D4$^d$ \cite{b97mv}                                                          & 5.68 & 5.44 & 37 & 0 & 1.95 & MB16-43   &  2 & 1 & 18.24 & MB16-43   \\
B97M-D3(BJ)$^c$ \cite{b97mv}                                                      & 6.44 & 6.41 & 33 & 3 & 2.81 & AL2X6     &  3 & 0 &  4.97 & ALK8      \\
r$^2$SCAN-D3(BJ)$^e$ \cite{r2scan}                                                & 7.12 & 6.79 & 25 & 0 & 1.72 & WCPT18    &  6 & 0 & 4.57  & SIE4x4    \\
B97M-D3(0)$^c$ \cite{b97mv}                                                       & 6.51 & 6.92 & 31 & 3 & 4.39 & ADIM6     &  5 & 1 & 28.23 & MB16-43   \\
revTPSS-D3(BJ)$^b$ \cite{revtpss}                                                 & 8.44 & 7.73 & 23 & 3 & 3.71 & SCONF     &  9 & 2 & 17.45 & MB16-43   \\
\hline
PW6B95-D3(BJ)$^{b,c}$ \cite{pw6b95}                                               & 5.50 & 5.14 & 41 & 1 & 2.31 & IL16      &  0 & 0 &  1.76 & SIE4x4    \\
M05-2X-D3(0)$^b$ \cite{m052x}                                                     & 4.62 & 5.19 & 41 & 1 & 2.08 & ADIM6     &  3 & 1 &  7.02 & MB16-43   \\
M06-2X-D3(0)$^b$ \cite{m062x}                                                     & 4.92 & 5.22 & 46 & 2 & 3.50 & HEAVYSB11 &  1 & 1 &  5.83 & HEAVYSB11 \\
PW6B95-V$^c$ \cite{pw6b95}                                                        & 5.57 & 5.24 & 45 & 2 & 2.49 & ADIM6     &  0 & 0 &  1.81 & SIE4x4    \\
revPBE0-D3(BJ)$^a$ \cite{zhang1993comment,perdew1996generalized,adamo1999toward}  & 5.31 & 5.30 & 43 & 0 & 1.51 & W4-11     &  1 & 0 &  3.03 & W4-11     \\
revPBE0-XDM(Z)$^a$ \cite{zhang1993comment,perdew1996generalized,adamo1999toward}  & 5.71 & 5.43 & 40 & 0 & 1.48 & W4-11     &  1 & 0 &  2.83 & W4-11     \\
revPBE0-XDM(BJ)$^a$ \cite{zhang1993comment,perdew1996generalized,adamo1999toward} & 5.06 & 5.44 & 42 & 1 & 2.58 & ALK8      &  2 & 1 &  7.71 & ALK8      \\
MPW1B95-D3(BJ)$^b$ \cite{mpw1,b95}                                                & 5.55 & 5.48 & 42 & 1 & 2.02 & ACONF     &  0 & 0 &  0.85 & RSE43     \\
B86bPBE0-XDM(BJ)$^a$ \cite{becke1986large,perdew1996generalized,adamo1999toward}  & 5.77 & 5.53 & 39 & 0 & 1.93 & ALK8      &  1 & 0 &  4.55 & ALK8      \\
M08-HX-D3(0)$^b$ \cite{m08hx}                                                     & 5.27 & 5.68 & 41 & 2 & 3.85 & ACONF     &  1 & 0 &  2.26 & C60ISO    \\
MPW1PW91-D3(BJ)$^b$ \cite{mpw1,pw91gga}                                           & 6.33 & 5.82 & 32 & 0 & 1.31 & PNICO23   &  0 & 0 &  1.04 & ALK8      \\
B86bPBE0-XDM(Z)$^a$ \cite{becke1986large,perdew1996generalized,adamo1999toward}   & 6.43 & 5.85 & 39 & 0 & 1.55 & SCONF     &  0 & 0 &  1.68 & WATER27   \\
\hline
$\omega$B97M-V$^{c,e}$ \cite{wb97mv}                                              & 3.18 & 3.72 & 49 & 1 & 2.22 & C60ISO    &  1 & 1 &  6.51 & C60ISO    \\
$\omega$B97M-D4$^d$ \cite{wb97mv}                                                 & 4.02 & 4.34 & 43 & 1 & 2.24 & C60ISO    &  2 & 2 &  6.60 & C60ISO    \\
$\omega$B97X-V$^{b,c}$ \cite{wb97xv}                                              & 3.93 & 4.40 & 50 & 1 & 2.57 & C60ISO    &  2 & 2 & 13.21 & MB16-43   \\
$\omega$B97M-D3(BJ)$^c$ \cite{wb97mv}                                             & 3.94 & 4.60 & 44 & 3 & 2.47 & C60ISO    &  2 & 1 &  7.86 & C60ISO    \\
$\omega$B97X-D4$^d$ \cite{wb97xv}                                                 & 4.33 & 4.68 & 46 & 1 & 2.65 & C60ISO    &  1 & 1 &  8.83 & C60ISO    \\
$\omega$B97M-D3(0)$^c$ \cite{wb97mv}                                              & 4.17 & 4.69 & 46 & 1 & 2.31 & C60ISO    &  2 & 1 &  7.01 & C60ISO    \\
$\omega$B97X-D3(BJ)$^c$ \cite{wb97xv}                                             & 4.35 & 5.27 & 46 & 2 & 2.99 & C60ISO    &  3 & 1 & 10.65 & C60ISO    \\
$\omega$B97X-D3(0)$^{b,c}$ \cite{wb97xd3}                                         & 4.73 & 5.38 & 40 & 1 & 2.54 & C60ISO    &  2 & 2 & 17.20 & MB16-43   \\
LC-$\omega$hPBE-XDM(BJ)$^{a,g}$ \cite{vydrov2006assessment, vydrov2007tests}      & 5.32 & 5.65 & 42 & 0 & 2.00 & ALK8      &  2 & 0 &  4.86 & ALK8      \\
HSE06-D3(BJ)$^b$ \cite{krukau2006influence}                                       & 6.83 & 6.08 & 32 & 0 & 1.43 & PCONF21   &  0 & 0 &  1.85 & WATER27   \\
LC-$\omega$hPBE-XDM(Z)$^{a,g}$ \cite{vydrov2006assessment, vydrov2007tests}       & 5.99 & 6.09 & 34 & 0 & 2.00 & SCONF     &  2 & 0 &  4.59 & C60ISO    \\
LC-$\omega$PBE-XDM(BJ)$^{a,h}$ \cite{vydrov2006assessment, vydrov2007tests}       & 5.31 & 6.39 & 32 & 2 & 3.17 & C60ISO    &  4 & 2 & 11.61 & C60ISO    \\
\hline
DH24$^b$ \cite{becke2024remarkably}                                               & 1.72 & 2.23 & 52 & 0 & 1.57 & C60ISO    &  1 & 0 & 3.07 & C60ISO     \\
revDH23$^b$ \cite{dh23}                                                           & 1.72 & 2.26 & 52 & 0 & 1.56 & C60ISO    &  1 & 0 & 2.99 & C60ISO     \\
SOS-DH24$^b$ \cite{becke2024remarkably}                                           & 1.91 & 2.27 & 52 & 0 & 1.26 & IL16      &  0 & 0 & 0.66 & C60ISO     \\
SOS-DH23$^b$ \cite{dh23}                                                          & 1.95 & 2.34 & 52 & 0 & 1.35 & IL16      &  0 & 0 & 0.73 & C60ISO     \\
$\omega$DOD-PBEP86-D3(BJ)$^{b,i}$ \cite{santra2021exploring}                      & 2.20 & 2.50 & 53 & 0 & 1.35 & ADIM6     &  0 & 0 & 0.73 & DIPCS10    \\
$\omega$DOD-PBEP86-D3(BJ)$^{b,j}$ \cite{santra2021exploring}                      & 2.21 & 2.52 & 53 & 0 & 1.54 & ADIM6     &  0 & 0 & 0.30 & DIPCS10    \\
\hline
DM21mu$^f$ \cite{kirkpatrick2021pushing}                                          & 3.93 & 3.72 & 50 & 1 & 2.91 & RG18      &  0 & 0 &  1.07 & HEAVYSB11 \\
DM21$^f$ \cite{kirkpatrick2021pushing}                                            & 3.97 & 3.94 & 48 & 2 & 2.68 & RG18      &  1 & 1 &  6.01 & C60ISO    \\
DM21mc$^f$ \cite{kirkpatrick2021pushing}                                          & 3.96 & 3.98 & 49 & 1 & 2.12 & ACONF     &  0 & 0 &  1.25 & C60ISO    \\
Skala$^e$ \cite{luise2025accurate}                                                & 3.83 & 4.01 & 47 & 0 & 1.60 & C60ISO    &  1 & 0 &  3.20 & C60ISO    \\
DM21m$^f$ \cite{kirkpatrick2021pushing}                                           & 3.89 & 4.04 & 48 & 1 & 2.98 & ACONF     &  0 & 0 &  0.82 & HEAVYSB11 \\
\hline
\end{tabular}\\
}
$^a$ Present work; 
$^b$ Ref.~\citenum{bryenton2026wtmad4} and references therein; 
$^c$ Ref.~\citenum{najibi2018nonlocal}; 
$^d$ Ref.~\citenum{najibi2020dft}; 
$^e$ Ref.~\citenum{luise2025accurate}; 
$^f$ Ref.~\citenum{kirkpatrick2021pushing}; \\
$^g$ $\omega=0.2$, $a_\text{X}=0.2$; 
$^h$ $\omega=0.4$; 
$^i$ $\omega=0.10$, $a_X=0.69$; 
$^j$ $\omega=0.08$, $a_X=0.72$\\
\end{table*}

Of the global hybrid functionals, revPBE0-D3(BJ), revPBE0-XDM(Z), MPW1B95-D3(BJ),\cite{mpw1,b95} MPW1PW91-D3(BJ),\cite{mpw1,pw91gga} and B86bPBE0-XDM(Z) all offer a good balance between a low \mbox{WTMAD-4} and few or no large outliers. MPW1B95-D3(BJ) involves the B95 meta-GGA correlation functional,\cite{b95} which is known to suffer from numerical instabilities.\cite{cpl,oscillations} However, the other four methods listed above use only GGA ingredients in their base functionals and appear to be the most consistently reliable choices. Each of the revPBE, MPW1, and B86b exchange functionals give good agreement with exact exchange repulsion in noble-gas dimers,\cite{zhang1993comment,mpw1,zhang1997describing} emphasising the advantages of dispersionless exchange in functional development. We note that the pairing of MPW1PW91 with XDM(Z) was not considered in this work as that functional is not implemented in FHI-aims (except via libxc\cite{lehtola2018recent}), but that may be a promising combination for future work.

In addition to our XDM variants, we also evaluated the performance of the TS, MBD@rsSCS, and \mbox{MBD-NL} dispersion corrections, all paired with PBE and PBE0. We note that MBD@rsSCS and MBD-NL failed for some systems due to a polarisation catastrophe.\cite{bryenton2023many} For MBDrsSCS, two reactions from ALK8 were substituted using MBD-NL results for each of PBE ($\ce{Na8} \rightarrow 4\,\ce{Na2}$ and $\ce{Li5CH} \rightarrow \ce{Li4C + LiH}$) and PBE0 ($\ce{Na8} \rightarrow 4\,\ce{Na2}$ and $\ce{(Li(CH)2N)2} \rightarrow 2 \, \ce{Li(CH)2N}$). Similarly, for MBD-NL, the forward and reverse barrier heights for the $\ce{F + H2} \rightarrow \ce{HF + H}$ reaction from BH76 were substituted using MBD@rsSCS results for both PBE and PBE0.

Ultimately, the GMTKN55 results for TS, MBD@rsSCS, and MBD-NL did not place among the best-ranking functional--DC combinations shown in Table~\ref{tab:rankings}, with WTMAD-4 values of 9.85, 9.44, and 9.42 for PBE, and 6.72, 6.30, and 6.36 for PBE0, respectively. Thus, we will limit our discussion to qualitative metrics, and the full data will be provided in the ESI. The TS method shows its largest errors for isomerisation energies and large systems (iso+large) category and, for the ALK8 set, MBD@rsSCS shows similar errors to XDM(BJ). Generally, MBD-NL and XDM(Z) are more consistently accurate across all GMTKN55 categories. Inspecting the outliers shows that PBE0-MBD-NL has $N_{r>2} = N_{d>2} = 0$, while also exhibiting good performance for MB16-43 with a MAE of 14.86. However, due to the aforementioned convergence issues and very limited choice of base functionals, we do not recommend these MBD methods for general thermochemical applications to molecular systems.

The analysis of outliers provides a much more nuanced assessment of the functionals than ranking by WTMAD-4 alone. The maximum difference outliers in particular show some distinct trends. For 4/6 top-ranked GGA functionals, the SIE4x4 set is the greatest outlier, as may be expected due to the inherent delocalisation error seen with this class of functional. Also, with the exception of r$^2$SCAN-D3(BJ),\cite{r2scan} the meta-GGAs tend to give either larger MAX $r$ or MAX $d$ outliers than the best-performing GGAs, making them difficult to recommend in most cases. Notably, the highly empirical B97M-D3(0)\cite{b97mv} gives a massive MAD of 47.53 kcal/mol for the MB16-43 set, indicating this functional is likely overfit as it provides unphysical performance for a benchmark that is chemically distinct from its training data. However, if one is willing to accept MAX d values of 16.67 or 18.24 kcal/mol for MB16-43 (corresponding to MAEs of  35.97 and 37.54 kcal/mol), B97M-V and B97M-D4\cite{b97mv} have far fewer outliers than the other meta-GGAs and GGAs, and \mbox{WTMAD-4} values comparable to the best hybrid functionals.

Turning to the hybrid functionals, it appears that low \mbox{WTMAD-4} values are often obtained at the expense of one or two benchmarks that are large outliers. This is quite evident for the more highly empirical functionals, such as the $\omega$B97 family. These $\omega$B97 functionals have low WTMAD-4 metrics, but tend to show large errors for the MB16-43 set (shown in the ESI). With the exception of $\omega$B97M-V,\cite{wb97mv} $\omega$B97X-D3BJ,\cite{wb97xv} and $\omega$B97X-D4\cite{wb97xv} all other variants have MB16-43 as a $d>2$ outlier with MADs of $>$20 kcal/mol. Particularly large outliers are seen for $\omega$B97X-D3(0)\cite{wb97xd3} and $\omega$B97X-V,\cite{wb97xv} again illustrating the problems with overfitting of highly empirical functionals. This family of functionals also reports high error outliers for the C60ISO benchmark. This C60ISO set, in general, remains a large outlier for other RS hybrids with full long-range exact exchange, which provides a poor description of large systems with highly extended conjugation and small band gaps. Correlation models with improved long-range physics are needed to pair well with exact (Hartree-Fock) exchange.\cite{becke2003real}

Finally, we extend our analysis to two machine-learned functionals that have garnered much recent attention: Skala,\cite{luise2025accurate} introduced by Microsoft in 2025; and DM21,\cite{kirkpatrick2021pushing} introduced by Google in 2021. We note that both of these functionals implicitly use D3(BJ) to account for London dispersion. Skala uses spin-indexed meta-GGA functional parameters as its features, specifically the electron density, its gradient, and the kinetic-energy density. Additionally, using a non-local interactions module, a course grid exchanges information between distant points, $r$ and $r^\prime$, capturing non-local information without resorting to the expensive four-centre two-electron Coulomb integrals required by hybrid functionals. With a WTMAD-4 of only 4.01, Skala's performance on GMTKN55 is markedly impressive. It only has one outlier of note, with a MAE for C60ISO of 8.54 kcal/mol (3.20 kcal/mol more than the $\overline{\text{MAD}_i}^\text{10-DFA}$ value), perhaps indicating a limitation of the non-local interactions module not being able to capture the physics of highly extended conjugation. DM21, meanwhile, has input features of the spin-indexed density, gradient, kinetic-energy density, and the range-separatedm exact-exchange energy density, making it a machine-learned, range-separated hybrid functional. DM21 was trained using constraints on fractional charge and spin, and three variants were created under different constraints: DM21m was unconstrained, DM21mc was constrained only on fractional charge, and DM21mu was constrained on the uniform electron gas (UEG) limit. Despite the additional input features, DM21 (and its variants) give comparable WTMAD-4 values to Skala. However, all have $N_{r>2}\ge1$, with maximum ratio outliers for the RG18 and ACONF sets that are dominated by London dispersion binding. Interestingly, DM21 also has a large MAE for C60ISO (11.35 kcal/mol) that is not present for its variants, suggesting this is due to the training constraints on fractional spin. Compared to Skala, DM21 has had more time for testing and, confirming initial suspicions,\cite{perdew2021artificial} DM21 was shown to have difficulties extrapolating beyond its training and validation sets.\cite{zhao2024deep} Ultimately, Skala was also designed with GMTKN55 as a validation set, and it will, thus, be interesting to see if Skala proves to be more robust and transferable than DM21.

\subsection{Molecular-Crystal Benchmarks}\label{ss:solids}

\begin{table*}[ht!]
\caption{
Mean absolute errors, in kcal/mol, for the X23, HalCrys4, and ICE13 (absolute and relative) lattice-energy benchmarks. All results are shown for \texttt{tight} basis settings at \texttt{lightdenser} geometries; for the hybrid functionals, this involved the basis-set correction of Eq.~\ref{eq:bsc}. \label{tab:molcrys}}
\centering
\begin{tabular}{l|cc|cc|cc|cc}\hline
 &         \multicolumn{2}{c|}{X23} & \multicolumn{2}{c|}{HalCrys4} & \multicolumn{2}{c|}{ICE13-Abs} & \multicolumn{2}{c}{ICE13-Rel} \\
Functional &BJ   &  Z   & BJ   & Z    & BJ   & Z    & BJ   & Z    \\ \hline
PBE        & 1.31 & 0.92 & 5.49 & 4.12 & 1.44 & 2.10 & 0.82 & 0.61 \\
revPBE     & 1.27 & 1.30 & 3.55 & 4.69 & 0.30 & 0.30 & 0.39 & 0.30 \\
B86bPBE    & 0.70 & 0.81 & 4.70 & 5.03 & 1.56 & 1.88 & 0.52 & 0.41 \\ \hline
PBE0       & 1.00 & 0.66 & 1.61 & 0.57 & 0.43 & 0.50 & 0.48 & 0.29 \\
revPBE0    & 0.90 & 0.71 & 2.47 & 1.12 & 1.36 & 0.79 & 0.23 & 0.21 \\
B86bPBE0   & 0.48 & 0.61 & 1.21 & 0.86 & 0.30 & 0.36 & 0.31 & 0.17 \\ \hline
PBE50      & 0.87 & 0.75 & 1.78 & 3.78 & 1.30 & 0.69 & 0.21 & 0.24 \\
revPBE50   & 0.72 & 0.74 & 2.81 & 3.71 & 2.18 & 1.41 & 0.18 & 0.39 \\
B86bPBE50  & 0.51 & 0.73 & 1.10 & 3.36 & 1.25 & 0.73 & 0.18 & 0.33 \\ \hline
\end{tabular}
\end{table*}

While XDM(Z) appears consistently reliable across the GMTKN55, it is also crucial to examine its performance for the solid state. Therefore, we consider the absolute lattice energies of the X23, HalCrys4, and ICE13 data sets, as well as the relative lattice energies of ICE13. Tabulated results for XDM(BJ) and XDM(Z) using the basis-set correction of Eq.~\ref{eq:bsc} are presented in Table~\ref{tab:molcrys}. 

The results in Table~\ref{tab:molcrys} show that the GGA functionals perform reasonably well, except for HalCrys4, where they overbind substantially due to delocalisation error. Similar overbinding is also seen for ICE13-Abs with XDM-corrected PBE and B86bPBE, but not revPBE, indicating an interplay between delocalisation error and the exchange enhancement factor. While anomalous for GGAs, the excellent performance of dispersion-corrected revPBE for both ICE13 benchmarks has been noted previously, leading to its popularity for simulations of water and ice.\cite{della2022dmc,nanowater} With the single exception of revPBE for ICE13-Abs, the 25\% hybrid functionals perform significantly better than their GGA counterparts, while further increases in exact-exchange mixing result in larger errors in most cases. Which of XDM(BJ) versus XDM(Z) is more accurate is highly dependent on both the benchmark and base functional. Nonetheless, XDM(Z) provides consistently good performance when paired with any of PBE0, revPBE0, or B86bPBE0. 

Only large-basis results have been considered in the above discussion to avoid confounding variables such as error cancellation. However, as shown in the ESI, the various methods also perform with exceptional accuracy and consistency for the molecular crystal benchmarks with the \texttt{lightdenser} basis setting, rivalling or even exceeding the basis-set-corrected results in Table~\ref{tab:molcrys}. This performance is worth noting, as these benchmarks are indicative of a method's effectiveness for crystal structure prediction (CSP). In CSP workflows, basis-set corrections are often used only for final energy refinement due to time and computational constraints; in practice, geometry optimisations and preliminary energy ranking typically employ a smaller basis such as \texttt{lightdenser}.

\section{Summary}

This work considers a new variant of the XDM dispersion model that addresses previous overbinding of metal clusters. It is the first study to test the XDM (and MBD) methods for the GMTKN55 data set, enabling a direct, head-to-head comparison of the most widely used dispersion corrections on a comprehensive benchmark for general main-group thermochemistry, kinetics, and non-covalent interactions. 
The canonical XDM(BJ) method showed strong results in all cases with the exception of the ALK8 benchmark, which originally motivated the study into Z damping.  XDM(Z) completely resolved this error and, despite eliminating one empirical parameter, still performs on par with other leading dispersion corrections for the  GMTKN55 set. We therefore recommend XDM(Z) as a good general method for both molecular and solid-state applications due to its consistent reliability. 

Overall, MPW1PW91-D3(BJ), B86bPBE0-XDM(Z), revPBE0-D3(BJ), and revPBE0-XDM(Z) are some of the best exchange-correlation functionals among those tested. Despite their simplicity, they give WTMAD-4 values only slightly higher than the leading hybrid functionals available in the literature, \cite{goerigk2017look,santra2019minimally} but with minimal outliers. revPBE0-XDM(Z) and revPBE0-D3(BJ) are particularly accurate for water clusters, while B86bPBE0-XDM(Z) and MPW1PW91-D3(BJ) are more accurate for atomisation energies. Any of these four methods is an excellent choice for a simple, minimally empirical density functional. 
For molecular crystals, XDM(Z) paired with any of PBE0, revPBE0, and B86bPBE0 demonstrates consistent accuracy. Thus, revPBE0-XDM(Z) and B86bPBE0-XDM(Z) emerge as reliable, minimally empirical methods that perform consistently well across molecular chemistry. PW6B95-D3(BJ) and PW6B95-V are also good choices as they give slightly lower WTMAD-4 values with minimal outliers, but have more complicated functional forms that rely on meta-GGA ingredients, which can result in numerical instabilities unless very large integration meshes are used.
Conversely, if one wishes to minimize WTMAD-4 at the expense of the functional having some outliers, B97M-V and B97M-D4 are good options for those that would tolerate large errors on MB16-43, which could transfer to uncommon/exotic systems. Skala, $\omega$B97M-V, and $\omega$B97X-D4 are good options if one would tolerate errors on C60ISO, which could affect large systems with extended conjugation more generally. 

Finally, analysis of the outliers (as opposite to only weighted mean absolute errors) was found to be particularly informative, and reveals weaknesses in particular DFAs that are not evident from their low WTMAD-4 values. It can be argued that introducing improved physics to eliminate the largest outliers is a better general strategy for ongoing functional development than introducing increasing numbers of empirical parameters to achieve slightly better across-the-board performance, which often comes at the expense of one or two larger outliers.

\section*{Acknowledgements}

KRB and ERJ thank the Natural Sciences and Engineering Research Council (NSERC) of Canada for financial support and the Atlantic Computing Excellence Network (ACENET) for computational resources. ERJ additionally thanks the Royal Society for a Wolfson Visiting Fellowship, while KRB thanks the Killam Trust, the Government of Nova Scotia, and the Mary Margaret Werner Graduate Scholarship Fund. Lastly, the authors thank A.\@ Simpson for permissions to use a picture of their pet gecko, Scamper, for the TOC image.

\section*{Data Availability Statement}

The data that support the findings of this study are available in the supplementary information.

\section*{Conflicts of Interest}

There are no conflicts to report.

\balance

\bibliographystyle{rsc}
\bibliography{main.bib}

@article{optx,
title={Left-right correlation energy},
author={Handy, N. C. and Cohen, A. J.},
journal={Mol. Phys.},
year={2001}, 
volume={99}, 
pages={403},
}

@article{r2scan,
title={Accurate and Numerically Efficient r$^2$SCAN Meta-Generalized Gradient Approximation},
author={James W. Furness and Aaron D. Kaplan and Jinliang Ning and John P. Perdew and Jianwei Sun},
journal={J. Phys. Chem. Lett.},
volume={11},
pages={8208-8215},
year={2020},
}

@article{luise2025accurate,
  title={Accurate and scalable exchange-correlation with deep learning},
  author={Giulia Luise and Chin-Wei Huang and Thijs Vogels and Derk P. Kooi and Sebastian Ehlert and Stephanie Lanius and Klaas J. H. Giesbertz and Amir Karton and Deniz Gunceler and Megan Stanley and Wessel P. Bruinsma and Lin Huang and Xinran Wei and Jos\'{e} Garrido Torres and Abylay Katbashev and Rodrigo Chavez Zavaleta and B\'{a}lint M\'{a}t\'{e} and S\'{e}kou-Oumar Kaba and Roberto Sordillo and Yingrong Chen and David B. Williams-Young and Christopher M. Bishop and Jan Hermann and Rianne van den Berg and Paola Gori-Giorgi},
  journal={arXiv preprint arXiv:2506.14665},
  year={2025},
  doi={10.48550/arXiv.2506.14665}
}

@article{revtpss,
author={John P. Perdew and Adrienn Ruzsinszky and Gábor I. Csonka and Lucian A. Constantin and Jianwei Sun}, 
title={Workhorse Semilocal Density Functional for Condensed Matter Physics and Quantum Chemistry}, 
journal={Phys. Rev. Lett.},
volume={103},
year={2009},
pages={026403},
}

@article{m08hx,
author={Y. Zhao and D. G. Truhlar}, 
title={Exploring the Limit of Accuracy of the Global Hybrid Meta Density Functional for Main-Group Thermochemistry, Kinetics, and Noncovalent Interactions}, 
journal={J. Chem. Theory Comput.},
year={2008}, 
volume={4}, 
pages={1849}
}

@article{pw91gga,
  title = {Atoms, molecules, solids, and surfaces: Applications of the generalized gradient approximation for exchange and correlation
},
 author = {Perdew, John P. and Chevary, J. A. and Vosko, S. H. and Jackson, Koblar A. and Pederson, Mark R. and Singh, D. J. and Fiolh
ais, Carlos},
  journal = {Phys. Rev. B},
  volume = {46},
  pages = {6671--6687},
  year = {1992},
}

@article{b95,
  author={A. D. Becke},
  title={Density-functional thermochemistry. 4. A new dynamical correlation functional and implications for exact-exchange mixing},
  journal={J. Chem. Phys.},
  volume={104},
  pages={1040-1046},
  year={1996},
}

@article{mpw1,
title={Exchange functionals with improved long-range behavior and adiabatic connection methods without adjustable parameters: The mPW and mPW1PW models},
author={Carlo Adamo and Vincenzo Barone},
journal={J. Chem. Phys.},
volume={108}, 
pages={664–675},
year={1998}
}

@article{pw6b95,
title={Design of Density Functionals That Are Broadly Accurate for Thermochemistry, Thermochemical Kinetics, and Nonbonded Interactions},
author={Yan Zhao and Donald G. Truhlar},
journal={J. Phys. Chem. A},
year={2005}, 
volume={109}, 
pages={5656-5667}
}

@article{m052x,
title={Design of Density Functionals by Combining the Method of Constraint Satisfaction with Parametrization for Thermochemistry, Thermochemical Kinetics, and Noncovalent Interactions},
author={Yan Zhao and Nathan E. Schultz and Donald G. Truhlar},
journal={J. Chem. Theory Comput.},
year={2006}, 
volume={2}, 
pages={364-382}
}

@article{m062x,
author={Zhao, Y and Truhlar, D G},
title={The M06 suite of density functionals for main group thermochemistry, thermochemical kinetics, noncovalent interactions, excited states, and transition elements: two new functionals and systematic testing of four M06-class functionals and 12 other functionals},
journal={Theor. Chem. Acc.},
year={2008},
volume={120},
pages={215–241} 
}

@article{b97mv,
title={Mapping the genome of meta-generalized gradient approximation density functionals: The search for B97M-V},
author={Narbe Mardirossian  and  Martin Head-Gordon},
journal={J. Chem. Phys.},
volume={142}, 
pages={074111},
year={2015}
}

@article{wb97mv,
title={$\omega$B97M-V: A combinatorially optimized, range-separated hybrid, meta-GGA density functional with VV10 nonlocal correlation},
author={Narbe Mardirossiana  and  Martin Head-Gordon},
journal={J. Chem. Phys.},
volume={144}, 
pages={214110},
year={2016}
}

@article{wb97xv,
title={$\omega$B97X-V: A 10-parameter, range-separated hybrid, generalized gradient approximation density functional with nonlocal correlation, designed by a survival-of-the-fittest strategy},
author={Narbe Mardirossiana  and  Martin Head-Gordon},
journal={Phys. Chem. Chem. Phys.}, 
year={2014},
volume={16}, 
pages={9904-9924}
}

@article{wb97xd3,
title={Long-Range Corrected Hybrid Density Functionals with Improved Dispersion Corrections},
author={You-Sheng Lin and Guan-De Li and Shan-Ping Mao and Jeng-Da Chai},
journal={J. Chem. Theory Comput.},
year={2013}, 
volume={9}, 
pages={263–272}
}

@article{lehtola2018recent,
  title={Recent developments in libxc—A comprehensive library of functionals for density functional theory},
  author={Lehtola, Susi and Steigemann, Conrad and Oliveira, Micael JT and Marques, Miguel AL},
  journal={SoftwareX},
  volume={7},
  pages={1--5},
  year={2018},
}

@article{zhang1993comment,
  title = {Comment on ``Generalized Gradient Approximation Made Simple''},
  author = {Zhang, Y. and Yang, W.},
  journal = {Phys. Rev. Lett.},
  volume = {80},
  year = {1998},
  pages = {890--890}
}

@article{nanowater,
title={The first-principles phase diagram of monolayer nanoconfined water},
author={Venkat Kapil and Christoph Schran and Andrea Zen and Ji Chen and Chris J. Pickard and Angelos Michaelides },
journal={Nature},
volume={609}, 
pages={512–516},
year={2022}
}

@article{elephant,
author={S. R. Jensen and S. Saha and J. A. Flores-Livas and W. Huhn and V. Blum and S. Goedecker and L. Frediani},
title={The elephant in the room of density functional theory calculations},
journal={J. Phys. Chem. Lett.}, 
volume={8},
pages={1449–1457}, 
year={2017},
}

@article{cpl,
title={Application of 25 density functionals to dispersion-bound homomolecular dimers},
author={Erin R. Johnson and Robert A. Wolkow and Gino A. DiLabio},
journal={Chem. Phys. Lett.},
volume={394},
pages={334-338},
year={2004},
}

@article{oscillations,
   title={Oscillations in meta-generalized-gradient approximation potential energy surfaces for dispersion-bound complexes},
   author={Johnson, Erin R. and Becke, Axel D. and Sherrill, C. David and DiLabio, Gino A.},
   journal={J. Chem. Phys.},
   volume={131},
   pages={034111},
   year={2009},
}

@article{wheeler,
title={Integration Grid Errors for Meta-GGA-Predicted Reaction Energies: Origin of Grid Errors for the M06 Suite of Functionals},
author={Steven E. Wheeler and K. N. Houk},
journal={J. Chem. Theory Comput.},
volume={6},
pages={395-404},
year={2010},
}

@article{headgordon,
title={How Accurate Are the Minnesota Density Functionals for Noncovalent Interactions, Isomerization Energies, Thermochemistry, and Barrier Heights Involving Molecules Composed of Main-Group Elements?},
author={Narbe Mardirossian and Martin Head-Gordon},
journal={J. Chem. Theory Comput.},
volume={12},
pages={4303-4325},
year={2016},
}

@article{bryenton2026wtmad4,
author={K. R. Bryenton and E. R. Johnson}, 
title={WTMAD-4: A Fair Weighting Scheme for GMTKN55},
journal={Phys. Chem. Chem. Phys.},
volume={28}, 
pages={1463-1469},
year={2026}
}

@article{becke2003real,
  title={A real-space model of nondynamical correlation},
  author={Becke, Axel D},
  journal={J. Chem. Phys.},
  volume={119},
  number={6},
  pages={2972--2977},
  year={2003},
  publisher={American Institute of Physics},
  doi={10.1063/1.1589733}
}

@article{becke2005density,
  title={A density-functional model of the dispersion interaction},
  author={Becke, Axel D and Johnson, Erin R},
  journal={J. Chem. Phys.},
  volume={123},
  number={15},
  pages={154101},
  year={2005},
  publisher={American Institute of Physics},
  doi={10.1063/1.2065267}
}

@article{christian2016surface,
  title={Surface adsorption from the exchange-hole dipole moment dispersion model},
  author={Christian, Matthew S and Otero-de-la-Roza, Alberto and Johnson, Erin R},
  journal={J. Chem. Theory Comput.},
  volume={12},
  number={7},
  pages={3305--3315},
  year={2016},
  publisher={ACS Publications},
  doi={10.1021/acs.jctc.6b00222}
}

@article{otero2014predicting,
  title={Predicting the Relative Solubilities of Racemic and Enantiopure Crystals by Density-Functional Theory},
  author={Otero-de-la-Roza, Alberto and Cao, Blessing Huynh and Price, Ivy K and Hein, Jason E and Johnson, Erin R},
  journal={Angew. Chem. Int. Ed.},
  volume={53},
  number={30},
  pages={7879--7882},
  year={2014},
  publisher={Wiley Online Library},
  doi={10.1002/ange.201403541}
}

@article{nickerson2023comparison,
  title={Comparison of density-functional theory dispersion corrections for the DES15K database},
  author={Nickerson, Cameron J and Bryenton, Kyle R and Price, Alastair JA and Johnson, Erin R},
  journal={J. Phys. Chem. A},
  volume={127},
  number={41},
  pages={8712--8722},
  year={2023},
  publisher={ACS Publications},
  doi={10.1021/acs.jpca.3c04332}
}

@article{price2023accurate,
author={A. J. A. Price and R. A. Mayo and A. {Otero-de-la-Roza} and E. R. Johnson}, 
title={Accurate and efficient polymorph energy ranking with XDM-corrected hybrid DFT},
journal={CrystEngComm},
volume={25}, 
pages={953-960},
year={2023}
}

@article{mayo2024assessment,
author={R. A. Mayo and A. J. A. Price and A. {Otero-de-la-Roza} and E. R. Johnson}, 
title={Assessment of the exchange-hole dipole moment dispersion correction for the energy ranking stage of the seventh crystal structure prediction blind test},
journal={Acta Crystallogr.},
volume={B80}, 
pages={595-605},
year={2024},
}

@article{otero2020application,
  title={Application of XDM to ionic solids: The importance of dispersion for bulk moduli and crystal geometries},
  author={Otero-de-la-Roza, Alberto and Johnson, Erin R},
  journal={J. Chem. Phys.},
  volume={153},
  number={5},
  pages={054121},
  year={2020},
  publisher={AIP Publishing LLC},
  doi={10.1063/5.0015133}
}

@article{perdew1996generalized,
  title={Generalized gradient approximation made simple},
  author={Perdew, John P and Burke, Kieron and Ernzerhof, Matthias},
  journal={Phys. Rev. Lett.},
  volume={77},
  number={18},
  pages={3865},
  year={1996},
  publisher={APS},
  doi={10.1103/PhysRevLett.77.3865}
}

@article{perdew1997erratum,
  title={Erratum: "Generalized gradient approximation made simple" (vol 77, pg 3865, 1996)},
  author={Perdew, John P and Burke, Kieron and Ernzerhof, Matthias},
  journal={Phys. Rev. Lett.},
  volume={78},
  number={7},
  pages={1396--1396},
  year={1997},
  publisher={APS},
  doi={10.1103/PhysRevLett.78.1396}
}

@article{adamo1999toward,
  title={Toward reliable density functional methods without adjustable parameters: The PBE0 model},
  author={Adamo, Carlo and Barone, Vincenzo},
  journal={J. Chem. Phys.},
  volume={110},
  number={13},
  pages={6158--6170},
  year={1999},
  publisher={American Institute of Physics},
  doi={10.1063/1.478522}
}

@article{price2021requirements,
  title={Requirements for an accurate dispersion-corrected density functional},
  author={Price, Alastair J A and Bryenton, Kyle R and Johnson, Erin R},
  journal={J. Chem. Phys.},
  volume={154},
  number={23},
  pages={230902},
  year={2021},
  publisher={AIP Publishing LLC},
  doi={10.1063/5.0050993}
}

@article{grimme2004accurate,
  title={Accurate description of van der Waals complexes by density functional theory including empirical corrections},
  author={Grimme, Stefan},
  journal={J. Comput. Chem.},
  volume={25},
  number={12},
  pages={1463--1473},
  year={2004},
  publisher={Wiley Online Library},
  doi={10.1002/jcc.20078}
}

@article{grimme2006semiempirical,
  title={Semiempirical GGA-type density functional constructed with a long-range dispersion correction},
  author={Grimme, Stefan},
  journal={J. Comput. Chem.},
  volume={27},
  number={15},
  pages={1787--1799},
  year={2006},
  publisher={Wiley Online Library},
  doi={10.1002/jcc.20495}
}

@article{grimme2010consistent,
  title={A consistent and accurate ab initio parametrization of density functional dispersion correction (DFT-D) for the 94 elements H-Pu},
  author={Grimme, Stefan and Antony, Jens and Ehrlich, Stephan and Krieg, Helge},
  journal={J. Chem. Phys.},
  volume={132},
  number={15},
  pages={154104},
  year={2010},
  publisher={American Institute of Physics},
  doi={10.1063/1.3382344}
}

@article{grimme2011effect,
  title={Effect of the damping function in dispersion corrected density functional theory},
  author={Grimme, Stefan and Ehrlich, Stephan and Goerigk, Lars},
  journal={J. Comput. Chem.},
  volume={32},
  number={7},
  pages={1456--1465},
  year={2011},
  publisher={Wiley Online Library},
  doi={10.1002/jcc.21759}
}

@article{caldeweyher2019generally,
  title={A generally applicable atomic-charge dependent London dispersion correction},
  author={Caldeweyher, Eike and Ehlert, Sebastian and Hansen, Andreas and Neugebauer, Hagen and Spicher, Sebastian and Bannwarth, Christoph and Grimme, Stefan},
  journal={J. Chem. Phys.},
  volume={150},
  number={15},
  pages={154122},
  year={2019},
  publisher={AIP Publishing LLC},
  doi={10.1063/1.5090222}
}

@inbook{johnson2017exchange,
  title={The Exchange-Hole Dipole Moment Dispersion Model},
  booktitle={Non-covalent Interactions in Quantum Chemistry and Physics: Theory and Applications},
  author={Johnson, Erin R},
  year={2017},
  publisher={Elsevier},
  chapter={5},
  pages={169--194},
  note={Edited by A. Otero-de-la-Roza and G. A. DiLabio},
  doi={10.1016/C2015-0-06383-3}
}

@article{johnson2006post,
  title={A post-Hartree-Fock model of intermolecular interactions: Inclusion of higher-order corrections},
  author={Johnson, Erin R and Becke, Axel D},
  journal={J. Chem. Phys.},
  volume={124},
  number={17},
  pages={174104},
  year={2006},
  publisher={American Institute of Physics},
  doi={10.1063/1.2190220}
}

@article{becke2007exchange,
  title={Exchange-hole dipole moment and the dispersion interaction revisited},
  author={Becke, Axel D and Johnson, Erin R},
  journal={J. Chem. Phys.},
  volume={127},
  number={15},
  pages={154108},
  year={2007},
  publisher={American Institute of Physics},
  doi={10.1063/1.2795701}
}

@article{christian2017adsorption2,
  title={Adsorption of Graphene to Metal (111) Surfaces using the Exchange-Hole Dipole Moment Model},
  author={Christian, Matthew S and Otero-de-la-Roza, Alberto and Johnson, Erin R},
  journal={Carbon},
  volume={124},
  pages={531-540},
  year={2017}
}

@article{adeleke2023effects,
author={A. A. Adeleke and E. R. Johnson}, 
title={Effects of dispersion corrections on the theoretical description of bulk metals},
journal={Phys. Rev. B},
volume={107}, 
pages={064101},
year={2023}
}

@article{otero2020asymptotic,
  title={Asymptotic Pairwise Dispersion Corrections Can Describe Layered Materials Accurately},
  author={Otero-de-la-Roza, Alberto and LeBlanc, Luc M and Johnson, Erin R},
  journal={J. Phys. Chem. Lett.},
  volume={11},
  number={6},
  pages={2298--2302},
  year={2020},
  publisher={ACS Publications},
  doi={10.1021/acs.jpclett.0c00348}
}

@article{becke1986large,
  title={On the large-gradient behavior of the density functional exchange energy},
  author={Becke, A D},
  journal={J. Chem. Phys.},
  volume={85},
  number={12},
  pages={7184--7187},
  year={1986},
  publisher={American Institute of Physics},
  doi={10.1063/1.451353}
}

@article{christian2021interplay,
  title={Interplay between London Dispersion, Hubbard U, and Metastable States for Uranium Compounds},
  author={Christian, Matthew S and Johnson, Erin R and Besmann, Theodore M},
  journal={J. Phys. Chem. A},
  volume={125},
  number={13},
  pages={2791--2799},
  year={2021},
  publisher={ACS Publications},
  doi={10.1021/acs.jpca.0c10533}
}

@article{dion2004van,
  title={Van der Waals density functional for general geometries},
  author={Dion, Max and Rydberg, Henrik and Schr{\"o}der, Elsebeth and Langreth, David C and Lundqvist, Bengt I},
  journal={Phys. Rev. Lett.},
  volume={92},
  number={24},
  pages={246401},
  year={2004},
  publisher={APS},
  doi={10.1103/PhysRevLett.92.246401}
}

@article{roman2009efficient,
  title={Efficient implementation of a van der Waals density functional: application to double-wall carbon nanotubes},
  author={Rom{\'a}n-P{\'e}rez, Guillermo and Soler, Jos{\'e} M},
  journal={Phys. Rev. Lett.},
  volume={103},
  number={9},
  pages={096102},
  year={2009},
  publisher={APS},
  doi={10.1103/PhysRevLett.103.096102}
}

@article{lee2010higher,
  title={Higher-accuracy van der Waals density functional},
  author={Lee, Kyuho and Murray, {\'E}amonn D and Kong, Lingzhu and Lundqvist, Bengt I and Langreth, David C},
  journal={Phys. Rev. B},
  volume={82},
  number={8},
  pages={081101},
  year={2010},
  publisher={APS},
  doi={10.1103/PhysRevB.82.081101}
}

@article{vydrov2010nonlocal,
  title={Nonlocal van der Waals density functional: The simpler the better},
  author={Vydrov, Oleg A and Van Voorhis, Troy},
  journal={J. Chem. Phys.},
  volume={133},
  number={24},
  pages={244103},
  year={2010},
  publisher={American Institute of Physics},
  doi={10.1063/1.3521275}
}

@article{sabatini2013nonlocal,
  title={Nonlocal van der Waals density functional made simple and efficient},
  author={Sabatini, Riccardo and Gorni, Tommaso and De Gironcoli, Stefano},
  journal={Phys. Rev. B},
  volume={87},
  number={4},
  pages={041108},
  year={2013},
  publisher={APS},
  doi={10.1103/PhysRevB.87.041108}
}

@article{reilly2013understanding,
  title={Understanding the role of vibrations, exact exchange, and many-body van der Waals interactions in the cohesive properties of molecular crystals},
  author={Reilly, Anthony M and Tkatchenko, Alexandre},
  journal={J. Chem. Phys.},
  volume={139},
  number={2},
  pages={024705},
  year={2013},
  publisher={American Institute of Physics},
  doi={10.1063/1.4812819}
}

@article{dolgonos2019revised,
  title={Revised values for the X23 benchmark set of molecular crystals},
  author={Dolgonos, Grygoriy A and Hoja, Johannes and Boese, A Daniel},
  journal={Phys. Chem. Chem. Phys.},
  volume={21},
  number={44},
  pages={24333--24344},
  year={2019},
  publisher={Royal Society of Chemistry},
  doi={10.1039/C9CP04488D}
}

@article{otero2012benchmark,
  title={A benchmark for non-covalent interactions in solids},
  author={Otero-de-la-Roza, A and Johnson, Erin R},
  journal={J. Chem. Phys.},
  volume={137},
  number={5},
  pages={054103},
  year={2012},
  publisher={American Institute of Physics},
  doi={10.1063/1.4738961}
}

@article{brandenburg2015benchmarking,
  title={Benchmarking DFT and semiempirical methods on structures and lattice energies for ten ice polymorphs},
  author={Brandenburg, Jan Gerit and Maas, Tilo and Grimme, Stefan},
  journal={J. Chem. Phys.},
  volume={142},
  number={12},
  pages={124104},
  year={2015},
  publisher={AIP Publishing},
  doi={10.1063/1.4916070}
}

@article{della2022dmc,
  title={DMC-ICE13: Ambient and high pressure polymorphs of ice from diffusion Monte Carlo and density functional theory},
  author={Della Pia, Flaviano and Zen, Andrea and Alf{\`e}, Dario and Michaelides, Angelos},
  journal={J. Chem. Phys.},
  volume={157},
  number={13},
  pages={134701},
  year={2022},
  publisher={AIP Publishing},
  doi={10.1063/5.0102645}
}

@article{otero2019dispersion,
  title={Dispersion XDM with hybrid functionals: Delocalization error and halogen bonding in molecular crystals},
  author={Otero-de-la-Roza, Alberto and LeBlanc, Luc M and Johnson, Erin R},
  journal={J. Chem. Theory Comput.},
  volume={15},
  number={9},
  pages={4933--4944},
  year={2019},
  publisher={ACS Publications},
  doi={10.1021/acs.jctc.9b00550}
}

@book{dean1999lange,
  title={Lange's Handbook of Chemistry},
  edition={15e},
  author={Dean, John A},
  year={1999},
  publisher={McGraw-Hill},
  isbn={978-0070163843}
}

@article{tkatchenko2009accurate,
  title={Accurate molecular van der Waals interactions from ground-state electron density and free-atom reference data},
  author={Tkatchenko, Alexandre and Scheffler, Matthias},
  journal={Phys. Rev. Lett.},
  volume={102},
  number={7},
  pages={073005},
  year={2009},
  publisher={APS},
  doi={10.1103/PhysRevLett.102.073005}
}

@article{tkatchenko2012accurate,
  title={Accurate and efficient method for many-body van der Waals interactions},
  author={Tkatchenko, Alexandre and DiStasio Jr, Robert A and Car, Roberto and Scheffler, Matthias},
  journal={Phys. Rev. Lett.},
  volume={108},
  number={23},
  pages={236402},
  year={2012},
  publisher={APS},
  doi={10.1103/PhysRevLett.108.236402}
}

@article{ambrosetti2014long,
  title={Long-range correlation energy calculated from coupled atomic response functions},
  author={Ambrosetti, Alberto and Reilly, Anthony M and DiStasio Jr, Robert A and Tkatchenko, Alexandre},
  journal={J. Chem. Phys.},
  volume={140},
  number={18},
  pages={18A508},
  year={2014},
  publisher={American Institute of Physics},
  doi={10.1063/1.4865104}
}

@article{gould2016fractionally,
  title={A fractionally ionic approach to polarizability and van der Waals many-body dispersion calculations},
  author={Gould, Tim and Lebegue, Sebastien and {\' A}ngy{\' a}n, J{\' a}nos G and Bu{\v c}ko Tom{\' a}{\v s}},
  journal={J. Chem. Theory Comput.},
  volume={12},
  number={12},
  pages={5920--5930},
  year={2016},
  publisher={ACS Publications},
  doi={10.1021/acs.jctc.6b00925}
}

@article{kim2020umbd,
  title={uMBD: A Materials-Ready Dispersion Correction That Uniformly Treats Metallic, Ionic, and van der Waals Bonding},
  author={Kim, Minho and Kim, Won June and Gould, Timothy and Lee, Eok Kyun and Lebegue, Sebastien and Kim, Hyungjun},
  journal={J. Am. Chem. Soc.},
  volume={142},
  number={5},
  pages={2346--2354},
  year={2020},
  publisher={ACS Publications},
  doi={10.1021/jacs.9b11589}
}

@article{hermann2020density,
  title={Density functional model for van der Waals interactions: unifying many-body atomic approaches with nonlocal functionals},
  author={Hermann, Jan and Tkatchenko, Alexandre},
  journal={Phys. Rev. Lett.},
  volume={124},
  number={14},
  pages={146401},
  year={2020},
  publisher={APS},
  doi={10.1103/PhysRevLett.124.146401}
}

@article{kannemann2010van,
  title={van der Waals interactions in density-functional theory: intermolecular complexes},
  author={Kannemann, Felix O and Becke, Axel D},
  journal={J. Chem. Theory Comput.},
  volume={6},
  number={4},
  pages={1081--1088},
  year={2010},
  publisher={ACS Publications},
  doi={10.1021/ct900699r}
}

@article{otero2013non,
  title={Non-covalent interactions and thermochemistry using XDM-corrected hybrid and range-separated hybrid density functionals},
  author={Otero-de-la-Roza, A and Johnson, Erin R},
  journal={J. Chem. Phys.},
  volume={138},
  number={20},
  pages={204109},
  year={2013},
  publisher={American Institute of Physics},
  doi={10.1063/1.4807330}
}

@inproceedings{perdew2001jacob,
  title={Jacob’s ladder of density functional approximations for the exchange-correlation energy},
  author={Perdew, John P and Schmidt, Karla},
  booktitle={AIP Conf. Proc.},
  volume={577},
  number={1},
  pages={1--20},
  year={2001},
  organization={American Institute of Physics},
  doi={10.1063/1.1390175}
}

@article{becke1988density,
  title={Density-functional exchange-energy approximation with correct asymptotic behavior},
  author={Becke, Axel D},
  journal={Phys. Rev. A},
  volume={38},
  number={6},
  pages={3098},
  year={1988},
  publisher={APS},
  doi={10.1103/PhysRevA.38.3098}
}

@article{becke1993new,
  title={A new mixing of Hartree-Fock and local density-functional theories},
  author={Becke, Axel D},
  journal={J. Chem. Phys.},
  volume={98},
  number={2},
  pages={1372--1377},
  year={1993},
  doi={10.1063/1.464304}
}

@article{becke1986density,
  title={Density functional calculations of molecular bond energies},
  author={Becke, Axel D},
  journal={J. Chem. Phys.},
  volume={84},
  number={8},
  pages={4524--4529},
  year={1986},
  publisher={American Institute of Physics},
  doi={10.1063/1.450025}
}

@article{beck1993density,
  title={Density-functional thermochemistry. III. The role of exact exchange},
  author={Beck, Axel D},
  journal={J. Chem. Phys},
  volume={98},
  number={7},
  pages={5648--5652},
  year={1993},
  doi={10.1063/1.464913}
}

@article{lee1988development,
  title={Development of the Colle-Salvetti correlation-energy formula into a functional of the electron density},
  author={Lee, Chengteh and Yang, Weitao and Parr, Robert G},
  journal={Phys. Rev. B},
  volume={37},
  number={2},
  pages={785},
  year={1988},
  publisher={APS},
  doi={10.1103/PhysRevB.37.785}
}

@article{stephens1994ab,
  title={Ab initio calculation of vibrational absorption and circular dichroism spectra using density functional force fields},
  author={Stephens, Philip J and Devlin, Frank J and Chabalowski, Cary F and Frisch, Michael J},
  journal={J. Phys. Chem.},
  volume={98},
  number={45},
  pages={11623--11627},
  year={1994},
  publisher={ACS Publications},
  doi={10.1021/j100096a001}
}

@article{vosko1980accurate,
  title={Accurate spin-dependent electron liquid correlation energies for local spin density calculations: a critical analysis},
  author={Vosko, Seymour H and Wilk, Leslie and Nusair, Marwan},
  journal={Can. J. Phys.},
  volume={58},
  number={8},
  pages={1200--1211},
  year={1980},
  publisher={NRC Research Press Ottawa, Canada},
  doi={10.1139/p80-159}
}

@article{vydrov2006assessment,
  title={Assessment of a long-range corrected hybrid functional},
  author={Vydrov, Oleg A and Scuseria, Gustavo E},
  journal={J. Chem. Phys.},
  volume={125},
  number={23},
  pages={234109},
  year={2006},
  publisher={AIP Publishing},
  doi={10.1063/1.2409292}
}

@article{vydrov2007tests,
  title={Tests of functionals for systems with fractional electron number},
  author={Vydrov, Oleg A and Scuseria, Gustavo E and Perdew, John P},
  journal={J. Chem. Phys.},
  volume={126},
  number={15},
  pages={154109},
  year={2007},
  publisher={AIP Publishing},
  doi={10.1063/1.2723119}
}

@article{krukau2006influence,
  title={Influence of the exchange screening parameter on the performance of screened hybrid functionals},
  author={Krukau, Aliaksandr V and Vydrov, Oleg A and Izmaylov, Artur F and Scuseria, Gustavo E},
  journal={J. Chem. Phys.},
  volume={125},
  number={22},
  pages={224106},
  year={2006},
  publisher={AIP Publishing},
  doi={10.1063/1.2404663}
}

@article{kirkpatrick2021pushing,
  title={Pushing the frontiers of density functionals by solving the fractional electron problem},
  author={Kirkpatrick, James and McMorrow, Brendan and Turban, David HP and Gaunt, Alexander L and Spencer, James S and Matthews, Alexander GDG and Obika, Annette and Thiry, Louis and Fortunato, Meire and Pfau, David and others},
  journal={Science},
  volume={374},
  number={6573},
  pages={1385--1389},
  year={2021},
  publisher={American Association for the Advancement of Science},
  doi={10.1126/science.abj6511}
}

@article{perdew2021artificial,
  title={Artificial intelligence “sees” split electrons},
  author={Perdew, John P},
  journal={Science},
  volume={374},
  number={6573},
  pages={1322--1323},
  year={2021},
  publisher={American Association for the Advancement of Science},
  doi={10.1126/science.abm2445}
}

@article{zhao2024deep,
  title={Deep Mind 21 functional does not extrapolate to transition metal chemistry},
  author={Zhao, Heng and Gould, Tim and Vuckovic, Stefan},
  journal={Physs. Chem. Chem. Phys.},
  volume={26},
  number={16},
  pages={12289--12298},
  year={2024},
  publisher={Royal Society of Chemistry},
  doi={10.1039/D4CP00878B}
}

@article{goerigk2017look,
  title={A look at the density functional theory zoo with the advanced {GMTKN55} database for general main group thermochemistry, kinetics and noncovalent interactions},
  author={Goerigk, Lars and Hansen, Andreas and Bauer, Christoph and Ehrlich, Stephan and Najibi, Asim and Grimme, Stefan},
  journal={Phys. Chem. Chem. Phys.},
  volume={19},
  number={48},
  pages={32184--32215},
  year={2017},
  publisher={Royal Society of Chemistry},
  doi={10.1039/C7CP04913G}
}

@article{santra2019minimally,
  title={Minimally empirical double-hybrid functionals trained against the {GMTKN55} database: revDSD-PBEP86-D4, revDOD-PBE-D4, and DOD-SCAN-D4},
  author={Santra, Golokesh and Sylvetsky, Nitai and Martin, Jan ML},
  journal={J. Phys. Chem. A},
  volume={123},
  number={24},
  pages={5129--5143},
  year={2019},
  publisher={ACS Publications},
  doi={10.1021/acs.jpca.9b03157}
}

@article{zhang1997describing,
  title={Describing van der Waals Interaction in diatomic molecules with generalized gradient approximations: The role of the exchange functional},
  author={Zhang, Yingkai and Pan, Wei and Yang, Weitao},
  journal={J. Chem. Phys.},
  volume={107},
  number={19},
  pages={7921--7925},
  year={1997},
  publisher={American Institute of Physics},
  doi={10.1063/1.475105}
}

@article{santra2021exploring,
  title={Exploring Avenues beyond Revised DSD Functionals: I. Range Separation, with x DSD as a Special Case},
  author={Santra, Golokesh and Cho, Minsik and Martin, Jan ML},
  journal={J. Phys. Chem. A},
  volume={125},
  number={21},
  pages={4614--4627},
  year={2021},
  publisher={ACS Publications},
  doi={10.1021/acs.jpca.1c01294}
}

@article{price2023xdm,
  title={XDM-corrected hybrid DFT with numerical atomic orbitals predicts molecular crystal lattice energies with unprecedented accuracy},
  author={Price, Alastair J A and {Otero-de-la-Roza}, Alberto and Johnson, Erin R},
  journal={Chem. Sci.},
  volume={14},
  number={5},
  pages={1252--1262},
  year={2023},
  publisher={Royal Society of Chemistry},
  doi={10.1039/D2SC05997E}
}

@article{bryenton2023many,
  title={Many-body dispersion in model systems and the sensitivity of self-consistent screening},
  author={Bryenton, Kyle R and Johnson, Erin R},
  journal={J. Chem. Phys.},
  volume={158},
  number={20},
  pages={204110},
  year={2023},
  publisher={AIP Publishing},
  doi={10.1063/5.0142465}
}

@article{blum2009ab,
  title={Ab initio molecular simulations with numeric atom-centered orbitals},
  author={Blum, Volker and Gehrke, Ralf and Hanke, Felix and Havu, Paula and Havu, Ville and Ren, Xinguo and Reuter, Karsten and Scheffler, Matthias},
  journal={Comput. Phys. Commun.},
  volume={180},
  number={11},
  pages={2175--2196},
  year={2009},
  publisher={Elsevier},
  doi={10.1016/j.cpc.2009.06.022}
}

@article{ren2012resolution,
  title={Resolution-of-identity approach to Hartree--Fock, hybrid density functionals, RPA, MP2 and GW with numeric atom-centered orbital basis functions},
  author={Ren, Xinguo and Rinke, Patrick and Blum, Volker and Wieferink, J{\"u}rgen and Tkatchenko, Alexandre and Sanfilippo, Andrea and Reuter, Karsten and Scheffler, Matthias},
  journal={New J. Phys.},
  volume={14},
  number={5},
  pages={053020},
  year={2012},
  publisher={IOP Publishing},
  doi={10.1088/1367-2630/14/5/053020}
}

@article{levchenko2015hybrid,
  title={Hybrid functionals for large periodic systems in an all-electron, numeric atom-centered basis framework},
  author={Levchenko, Sergey V and Ren, Xinguo and Wieferink, J{\"u}rgen and Johanni, Rainer and Rinke, Patrick and Blum, Volker and Scheffler, Matthias},
  journal={Comput. Phys. Commun.},
  volume={192},
  pages={60--69},
  year={2015},
  publisher={Elsevier},
  doi={10.1016/j.cpc.2015.02.021}
}

@article{kokott2024efficient,
  title={Efficient All-electron Hybrid Density Functionals for Atomistic Simulations Beyond 10,000 Atoms},
  author={Kokott, Sebastian and Merz, Florian and Yao, Yi and Carbogno, Christian and Rossi, Mariana and Havu, Ville and Rampp, Markus and Scheffler, Matthias and Blum, Volker},
  journal={J. Chem. Phys.},
  volume={161},
  number={2},
  pages={024112},
  year={2024},
  publisher={AIP Publishing},
  doi={10.1063/5.0208103}
}

@article{yu2018elsi,
  title={ELSI: A unified software interface for Kohn--Sham electronic structure solvers},
  author={Yu, Victor Wen-zhe and Corsetti, Fabiano and Garc{\'\i}a, Alberto and Huhn, William P and Jacquelin, Mathias and Jia, Weile and Lange, Bj{\"o}rn and Lin, Lin and Lu, Jianfeng and Mi, Wenhui and Ali Seifitokaldani and \'{A}. V\'{a}zquez-Mayagoitia and Chao Yang and Haizhao Yang and Volker Blum},
  journal={Comput. Phys. Commun.},
  volume={222},
  pages={267--285},
  year={2018},
  publisher={Elsevier},
  doi={10.1016/j.cpc.2017.09.007}
}

@article{havu2009efficient,
  title={Efficient O (N) integration for all-electron electronic structure calculation using numeric basis functions},
  author={Havu, Ville and Blum, Volker and Havu, Paula and Scheffler, Matthias},
  journal={J. Chem. Phys.},
  volume={228},
  number={22},
  pages={8367--8379},
  year={2009},
  publisher={Elsevier},
  doi={doi.org/10.1016/j.jcp.2009.08.008}
}

@article{ihrig2015accurate,
  title={Accurate localized resolution of identity approach for linear-scaling hybrid density functionals and for many-body perturbation theory},
  author={Ihrig, Arvid Conrad and Wieferink, J{\"u}rgen and Zhang, Igor Ying and Ropo, Matti and Ren, Xinguo and Rinke, Patrick and Scheffler, Matthias and Blum, Volker},
  journal={New J. Phys.},
  volume={17},
  number={9},
  pages={093020},
  year={2015},
  publisher={IOP Publishing},
  doi={10.1088/1367-2630/17/9/093020}
}

@article{becke2024remarkably,
  title={A remarkably simple dispersion damping scheme and the DH24 double hybrid density functional},
  author={Becke, Axel D},
  journal={J. Chem. Phys.},
  volume={160},
  number={20},
  pages={204118},
  year={2024},
  publisher={AIP Publishing},
  doi={10.1063/5.0207682}
}

@article{burke2016locality,
  title={Locality of correlation in density functional theory},
  author={Burke, Kieron and Cancio, Antonio and Gould, Tim and Pittalis, Stefano},
  journal={J. Chem. Phys.},
  volume={145},
  number={5},
  year={2016},
  pages={054112},
  publisher={AIP Publishing},
  doi={10.1063/1.4959126}
}

@article{dh23,
title={Doubling down on density-functional theory},
author={Axel D. Becke},
journal={J. Chem. Phys.},
volume={159}, 
pages={241101},
year={2023}
}

@software{otero2015refdata,
  author={{Otero-de-la-Roza}, Alberto},
  title={refdata},
  month={Sep},
  year={2015},
  publisher={GitHub},
  version={2025-02-04},
  url={https://github.com/aoterodelaroza/refdata}
}

@software{johnson2024gmtkn55,
  author={Johnson, Erin R},
  title={gmtkn55-fhiaims},
  month={Dec},
  year={2024},
  publisher={GitHub},
  version={2024-12-23},
  url={https://github.com/erin-r-johnson/gmtkn55-fhiaims}
}

@software{otero2025postgxcdm,
  author={{Otero-de-la-Roza}, Alberto and Bryenton, Kyle R. and Kannemann, Felix and Johnson, Erin R. and Dickson, Ross M. and Schmider, Harmut and Becke, Axel D.},
  title={postg (release: {XCDM(Z)})},
  month={Sep},
  year={2015},
  publisher={GitHub},
  version={2025-05-23},
  url={https://github.com/aoterodelaroza/postg}
}

@article{wittmann2023dispersion,
  title={Dispersion-corrected r2SCAN based double-hybrid functionals},
  author={Wittmann, Lukas and Neugebauer, Hagen and Grimme, Stefan and Bursch, Markus},
  journal={J. Chem. Phys.},
  volume={159},
  number={22},
  pages={224103},
  year={2023},
  publisher={AIP Publishing},
  doi={10.1063/5.0174988}
}

@article{bryenton2023delocalization,
  title={Delocalization error: The greatest outstanding challenge in density-functional theory},
  author={Bryenton, Kyle R and Adeleke, Adebayo A and Dale, Stephen G and Johnson, Erin R},
  journal={WIRES: Comput. Mol. Sci.},
  volume={13},
  number={2},
  pages={e1631},
  year={2023},
  publisher={Wiley Online Library},
  doi={10.1002/wcms.1631}
}

@article{hoja2018first,
  title={First-principles stability ranking of molecular crystal polymorphs with the DFT+MBD approach},
  author={J. Hoja and A. Tkatchenko},
  journal={Faraday Discuss.},
  volume={211},
  pages={253-274},
  year={2018},
}

@article{abbott2025roadmap,
  title={Roadmap on Advancements of the FHI-aims Software Package},
  author={Joseph W. Abbott and Carlos Mera Acosta and Alaa Akkoush and Alberto Ambrosetti and Viktor Atalla and Alexej Bagrets and J\"{o}rg Behler and Daniel Berger and Bj\"{o}rn Bieniek and Jonas Bj\"{o}rk and Volker Blum and Saeed Bohloul and Connor L. Box and Nicholas Boyer and Danilo Simoes Brambila and Gabriel A. Bramley and Kyle R. Bryenton and Mar\'{i}a Camarasa-G\'{o}mez and Christian Carbogno and Fabio Caruso and Sucismita Chutia and Michele Ceriotti and G\'{a}bor Cs\'{a}nyi and William Dawson and Francisco A. Delesma and Fabio Della Sala and Bernard Delley and Robert A. DiStasio Jr. and Maria Dragoumi and Sander Driessen and Marc Dvorak and Simon Erker and Ferdinand Evers and Eduardo Fabiano and Matthew R. Farrow and Florian Fiebig and Jakob Filser and Lucas Foppa and Lukas Gallandi and Alberto Garcia and Ralf Gehrke and Simiam Ghan and Luca M. Ghiringhelli and Mark Glass and Stefan Goedecker and Dorothea Golze and James A. Green and Andrea Grisafi and Andreas Gr\"{u}neis and Jan Günzl and Stefan Gutzeit and Samuel J. Hall and Felix Hanke and Ville Havu and Xingtao He and Joscha Hekele and Olle Hellman and Uthpala Herath and Jan Hermann and Daniel Hernang\'{o}mez-P\'{e}rez and Oliver T. Hofmann and Johannes Hoja and Simon Hollweger and Lukas H\"{o}rmann and Ben Hourahine and Wei Bin How and William P. Huhn and Marcel H\"{u}lsberg and Sara Panahian Jand and Hong Jiang and Erin R. Johnson and Werner Jürgens and J. Matthias Kahk and Yosuke Kanai and Kisung Kang and Petr Karpov and Elisabeth Keller and Roman Kempt and Danish Khan and Matthias Kick and Benedikt P. Klein and Jan Kloppenburg and Alexander Knoll and Florian Knoop and Franz Knuth and Simone S. K\"{o}cher and Jannis Kockl\"{a}uner and Sebastian Kokott and Thomas K\"{o}rzd\"{o}rfer and Hagen-Henrik Kowalski and Peter Kratzer and Pavel K{\r{u}}s and Raul Laasner and Bruno Lang and Bj\"{o}rn Lange and Marcel F. Langer and Ask Hjorth Larsen and Hermann Lederer and Susi Lehtola and Maja-Olivia Lenz-Himmer and Moritz Leucke and Sergey Levchenko and Alan Lewis and O. Anatole von Lilienfeld and Konstantin Lion and Werner Lipsunen and Johannes Lischner and Yair Litman and Chi Liu and Qing-Long Liu and Andrew J. Logsdail and Michael Lorke and Zekun Lou and Iuliia Mandzhieva and Andreas Marek and Johannes T. Margraf and Reinhard J. Maurer and Tobias Melson and Florian Merz and J\"{o}rg Meyer and Georg S. Michelitsch and Teruyasu Mizoguchi and Evgeny Moerman and Dylan Morgan and Jack Morgenstein and Jonathan Moussa and Akhil S. Nair and Lydia Nemec and Harald Oberhofer and Alberto {Otero-de-la-Roza} and Ram\'{o}n L. Panad\'{e}s-Barrueta and Thanush Patlolla and Mariia Pogodaeva and Alexander P\"{o}ppl and Alastair J. A. Price and Thomas A. R. Purcell and Jingkai Quan and Nathaniel Raimbault and Markus Rampp and Karsten Rasim and Ronald Redmer and Xinguo Ren and Karsten Reuter and Norina A. Richter and Stefan Ringe and Patrick Rinke and Simon P. Rittmeyer and Herzain I. Rivera-Arrieta and Matti Ropo and Mariana Rossi and Victor Ruiz and Nikita Rybin and Andrea Sanfilippo and Matthias Scheffler and Christoph Scheurer and Christoph Schober and Franziska Schubert and Tonghao Shen and Christopher Shepard and Honghui Shang and Kiyou Shibata and Andrei Sobolev and Ruyi Song and Aloysius Soon and Daniel T. Speckhard and Pavel V. Stishenko and Muhammad Tahir and Izumi Takahara and Jun Tang and Zechen Tang and Thomas Theis and Franziska Theiss and Alexandre Tkatchenko and Milica Todorovi\'{c} and George Trenins and Oliver T. Unke and \'{A}lvaro V\'{a}zquez-Mayagoitia and Oscar van Vuren and Daniel Waldschmidt and Han Wang and Yanyong Wang and J\"{u}rgen Wieferink and Jan Wilhelm and Scott Woodley and Jianhang Xu and Yong Xu and Yi Yao and Yingyu Yao and Mina Yoon and Victor Wen-zhe Yu and Zhenkun Yuan and Marios Zacharias and Igor Ying Zhang and Min-Ye Zhang and Wentao Zhang and Rundong Zhao and Shuo Zhao and Ruiyi Zhou and Yuanyuan Zhou and Tong Zhu},
  journal={arXiv preprint arXiv:2505.00125},
  year={2025},
  doi={10.48550/arXiv.2505.00125}
}

@article{najibi2018nonlocal,
  title={The nonlocal kernel in van der Waals density functionals as an additive correction: An extensive analysis with special emphasis on the B97M-V and $\omega$B97M-V approaches},
  author={Najibi, Asim and Goerigk, Lars},
  journal={J. Chem. Theory Comput.},
  volume={14},
  number={11},
  pages={5725--5738},
  year={2018},
  publisher={ACS Publications},
  doi={10.1021/acs.jctc.8b00842}
}

@article{najibi2020dft,
  title={DFT-D4 counterparts of leading meta-generalized-gradient approximation and hybrid density functionals for energetics and geometries},
  author={Najibi, Asim and Goerigk, Lars},
  journal={J. Comput. Chem.},
  volume={41},
  number={30},
  pages={2562--2572},
  year={2020},
  publisher={Wiley Online Library},
  doi={10.1002/jcc.26411}
}

@article{van1994relativistic,
  title={Relativistic total energy using regular approximations},
  author={van Lenthe, Erik and Baerends, Evert-Jan and Snijders, Jaap G},
  journal={J. Chem. Phys.},
  volume={101},
  number={11},
  pages={9783--9792},
  year={1994},
  publisher={American Institute of Physics},
  doi={10.1063/1.467943}
}

\end{document}